\documentclass[prd,superscriptaddress,twocolumn,showpacs,amsmath,%
preprintnumbers]{revtex4}
\usepackage{bm} 
\usepackage{graphicx}
\newcommand{\beq}{\begin{equation}}
\newcommand{\eeq}{\end{equation}}
\newcommand{\be}{\begin{eqnarray}}
\newcommand{\ee}{\end{eqnarray}}
\begin{document}
\author{L.~Frankfurt}
\affiliation{School of Physics and Astronomy, Tel Aviv University, 
Tel Aviv, Israel}
\author{C.E.~Hyde--Wright}
\affiliation{Old Dominion University, Norfolk, VA 23529, USA}
\author{M.~Strikman}
\affiliation{Department of Physics, Pennsylvania State University,
University Park, PA 16802, USA}
\author{C.~Weiss}
\affiliation{Theory Center, Jefferson Lab, Newport News, VA 23606, USA}
\title{Generalized parton distributions and rapidity gap survival \\
in exclusive diffractive $pp$ scattering}
\begin{abstract}
We propose a new approach to the problem of rapidity gap survival (RGS) 
in the production of high--mass systems ($H$ = dijet, heavy quarkonium,
Higgs boson) in double--gap exclusive diffractive $pp$ scattering, 
$pp \rightarrow p + \text{(gap)} + H + \text{(gap)} + p$.
It is based on the idea that hard and soft interactions proceed over 
widely different time-- and distance scales and are thus approximately 
independent. The high--mass system is produced in a hard scattering 
process with exchange of two gluons between the protons. Its amplitude 
is calculable in terms of the gluon generalized parton distributions (GPDs)
in the protons, which can be measured in $J/\psi$ production in exclusive 
$ep$ scattering. The hard scattering process is modified by soft spectator 
interactions, which we calculate in a model--independent way in terms of 
the $pp$ elastic scattering amplitude. Contributions from inelastic 
intermediate states are suppressed. A simple geometric picture of the 
interplay of hard and soft interactions in diffraction is obtained. 
The onset of the black--disk 
limit in $pp$ scattering at TeV energies strongly suppresses diffraction
at small impact parameters and is the main factor in determining the 
RGS probability. Correlations between hard and soft interactions
(\textit{e.g.}\ due to scattering from the long--range pion field of 
the proton, or due to possible short--range transverse correlations
between partons) further decrease the RGS probability.
We also investigate the dependence of the diffractive cross section on the 
transverse momenta of the final--state protons (``diffraction pattern'').
By measuring this dependence one can perform detailed tests of the
interplay of hard and soft interactions, and even extract information 
about the gluon GPD in the proton. Such studies appear to be feasible 
with the planned forward detectors at the LHC.
\end{abstract}
\preprint{JLAB-THY-06-533}
\maketitle
\section{Introduction}
Hard processes in high--energy $pp$ scattering are important both
as a laboratory for studying strong interaction dynamics and the
parton structure of the proton, and as one of the main tools 
in the search for new heavy particles. Of particular interest are 
so--called diffractive processes, in which the produced high--mass 
system (dijet, heavy particle) is separated from the projectile 
fragments by large rapidity gaps. Double--gap exclusive processes 
(\textit{i.e.}, without breakup of the protons)
\beq
pp \;\; \rightarrow \;\; p + \text{(gap)} + H + \text{(gap)} + p,
\label{exclusive_diffraction}
\eeq
are considered as an option for the Higgs boson search at the LHC
\cite{Schafer:1990fz,Bialas:1991wj,Dokshitzer:1991he,%
Khoze:2000wk,Kaidalov:2003fw}. 
Such processes have lower cross section than inclusive 
double--gap processes (with breakup of one or both protons), 
but offer better chances for detection, and for 
determining the mass of the produced particle and possibly even
its quantum numbers; see Ref.~\cite{Kaidalov:2003fw} and 
references therein. Double--gap exclusive processes 
(\ref{exclusive_diffraction}) also appear to be an effective method
for producing heavy quarkonia and investigating their properties.

From the point of view of strong interactions, double--gap exclusive 
events (\ref{exclusive_diffraction}) arise as the result of 
an interesting interplay of ``hard'' (involving momentum transfers 
much larger than the typical hadronic mass scale)
and ``soft'' (momentum transfers of the order of the typical hadronic 
mass scale) interactions. The high--mass system is produced in a 
hard scattering process, involving the exchange of two gluons
between the protons. The requirement of the absence of QCD radiation
ensures the localization of this process in space and time.
This alone, however, is not sufficient to guarantee a diffractive event.
One must also require that the soft interactions between
the spectator systems do not lead to particle production.
This results in a suppression of the cross section as compared to 
the hard scattering process alone, the so--called 
rapidity gap survival (RGS) probability. While not directly
observable, this quantity plays a central role in the discussion 
of hard diffractive processes and their use in new particle searches.

Diffractive final states are most favorably produced in scattering
at large impact parameters (peripheral scattering), where the chances
for the spectator systems not to interact inelastically are large.
Conversely, this means that the selection of diffractive events
changes the effective impact parameters as compared to inclusive
events with the same hard scattering process. This effect is 
essential for understanding the physical mechanism of RGS. In this sense, 
RGS is a manifestation of a general quantum--mechanical 
phenomenon --- the postselection of certain initial--state configurations 
by conditions imposed on the final state \cite{Nussinov}.

The concept of RGS should also be viewed in the context of QCD 
factorization for the production of heavy particles in 
$pp$ scattering \cite{Bjorken:1992er}. QCD factorization 
was formally proved for inclusive scattering, $pp \rightarrow H + X$.
A crucial element in this is the cancellation of initial--state
and final--state QCD radiation. This cancellation becomes incomplete 
if additional conditions, such as rapidity gaps, are imposed 
on the hadronic final state. The introduction of the RGS probability
can be seen as an attempt to ``restore'' QCD factorization for 
diffractive processes at the phenomenological level, 
in a form analogous to the inclusive case.

In this paper, we propose a new approach to RGS in double--gap exclusive 
hard diffractive processes. It is based on the idea that hard and soft 
interactions are approximately independent because they happen over widely 
different time-- and distance scales. We implement this idea in the
framework of a partonic description of the proton, along the lines
of Gribov's parton picture of high--energy hadron--hadron 
scattering \cite{Gribov:1973jg}. In the approximation where hard
and soft interactions are considered to be completely independent,
one can regard the hard scattering process as a ``local operator''
in partonic states, and RGS appears as the ``renormalization'' 
of this operator due to soft interactions. At the amplitude level, 
this leads to an absorption correction to the 
hard production process due to elastic rescattering;
contributions from inelastic intermediate states are suppressed
because of the different character of the states accessible in hard and 
soft interactions. At the cross section level, we recover 
a simple ``geometric'' expression for the RGS probability, which was 
suggested on the basis of heuristic arguments in 
Refs.~\cite{Frankfurt:2004kn,Frankfurt:2004ti,Frankfurt:2005mc}.
In addition to providing a transparent physical picture of RGS, 
this expression can readily be evaluated in terms of two phenomenological 
ingredients, both of which can be probed in independent measurements:
\begin{itemize}
\item 
The gluon generalized parton distribution (GPD) in the proton;
more precisely its $t$--dependence (``two--gluon formfactor''),
whose Fourier transform describes the transverse spatial 
distribution of gluons. Information about it
comes from measurements of hard exclusive processes in $ep$ scattering, 
in particular $J/\psi$ photoproduction (HERA, FNAL).
\item
The $pp$ elastic scattering amplitude at high energies; 
in particular its profile function in the impact parameter 
representation. It is known from fits to $pp/\bar pp$ total and
elastic cross section data up to the Tevatron energy, and constrained
by general arguments (black--disk limit, see below)
\end{itemize}

The framework provided by our partonic approach to RGS allows us 
to take into account two basic facts about the dynamics of hard 
and soft interactions at high energies, which turn out to have a 
decisive influence on the numerical value of the RGS probability. 
These are:
\begin{itemize}
\item \textit{Small transverse radius of hard interactions.} 
The radius of the transverse distribution of hard gluons in the proton 
is significantly smaller than the transverse radius of soft interactions 
in high--energy $pp$ collisions (``two--scale picture''). This 
basic fact explains many qualitative features of hard exclusive 
diffractive processes, such as the effective impact parameters
in diffractive events, the order--of--magnitude of the RGS probability,
and the pattern of the transverse momentum dependence of the cross section.
\item \textit{Black--disk limit in high--energy $pp$ scattering.}
Parametrizations of the data as well as general theoretical arguments
indicate that the profile function of $pp$ elastic scattering becomes
``black'' at small impact parameters at energies above the
Tevatron energy, $\sqrt{s} > 2\, \text{TeV}$. This circumstance
makes the description of $pp$ scattering at small impact parameters at
LHC energies practically model--independent, and is essential for the
stability of numerical predictions for the RGS.
\end{itemize}
The evidence supporting these statements is described in detail in 
Ref.~\cite{Frankfurt:2005mc}, and summarized in 
Sections~\ref{sec:transverse_soft} and \ref{sec:transverse_hard} below.

Our partonic approach also allows us to describe the effects of correlations 
between hard and soft interactions on the RGS probability. Such 
correlations can arise from various dynamical mechanisms,
\textit{e.g.}\ from the long--range pion field of the proton, 
or from possible short--range transverse correlations between 
hard partons, as suggested by the Tevatron CDF data on inclusive $pp$ 
scattering with multiple hard processes \cite{Abe:1997bp}.
We find that the inclusion of such correlations decreases the RGS 
probability compared to the independent interaction approximation.
While these effects cannot be calculated in a completely 
model--independent way, they are important both for our general 
understanding of the mechanism of RGS, and for obtaining reliable 
numerical estimates of the RGS probability. 
These effects clearly merit further study.

A unique feature of exclusive diffractive processes is that the
interplay of hard and soft interactions can be studied experimentally,
by measuring the dependence of the cross section on the transverse 
momenta of the final--state protons. The modification of the hard
scattering amplitude by soft elastic rescattering can be viewed
as an interference phenomenon, which gives rise to a distinctive
``diffraction pattern'' in the final--state transverse momenta.
By measuring this dependence in exclusive diffractive processes
with relatively large cross section, such as dijet production,
one can perform a variety of tests of the diffractive reaction mechanism, 
and extract information about the transverse radii of hard and soft 
interactions and their energy dependence. In Higgs production, measurements
of the transverse momentum dependence would allow one to obtain additional 
information about the parity of the produced particle \cite{Kaidalov:2003fw}. 
Experimentally, such studies appear to be feasible with the planned forward 
detectors at the LHC \cite{Albrow:2005ig} and the 
Tevatron \cite{Albrow:2005fw}. 

Detailed studies of RGS in diffractive $pp$ scattering were done 
in a model of soft interactions based on eikonalized Pomeron 
exchange \cite{Khoze:2000wk,Kaidalov:2003fw}. We show here
that, in the approximation where hard and soft interactions are 
considered to be independent, the RGS probability unambiguously 
follows from QCD and can be calculated in a model--independent way. 
Nevertheless, our numerical results for the RGS probability in this 
approximation turn out to be of the same order of magnitude as those 
reported in Refs.~\cite{Khoze:2000wk,Kaidalov:2003fw}, which can be 
attributed to the fact that the Pomeron parametrization of the $pp$ 
elastic amplitude reproduces the approach to the BDL at small impact 
parameters. The agreement between our numerical results and 
theirs at the quantitative level is somewhat accidental, being due to the 
fact that in the calculation of Refs.~\cite{Khoze:2000wk,Kaidalov:2003fw} 
the effect from inelastic intermediate states (which in our approach 
are seen to be strongly suppressed because of the small overlap of states 
accessible via hard and soft interactions) is partly compensated by 
the choice of a larger value of the $t$--slope of the gluon GPD; see 
Section~\ref{sec:cross_section} for details. Finally, our partonic 
approach can naturally be extended to include correlations between 
hard and soft interactions, which have a potentially large numerical
effect on the RGS probability.

This paper is organized as follows. In Section~\ref{sec:transverse_soft}
we review the information about the transverse structure of soft interactions 
from $pp$ elastic scattering, and the approach to the black--disk 
limit at central impact parameters. 
In Section~\ref{sec:transverse_hard} we discuss the properties
of the proton's gluon GPD at small $x$ and summarize our knowledge 
of the transverse spatial distribution of hard partons in the proton.
Section~\ref{sec:gap_survival} describes the basic framework of our 
approach to RGS. We outline the properties of the hard scattering 
amplitude, describe the theoretical formulation of the independence 
of hard and soft interactions, and obtain a master expression for 
the diffractive amplitude combining hard and soft interactions. 
We then explain the suppression of inelastic intermediate states, 
and evaluate the diffractive amplitude in terms of the gluon GPD 
and the $pp$ elastic amplitude. 
In Section~\ref{sec:cross_section} we use our result for the 
amplitude in the independent interaction approximation
to calculate the RGS probability. We recover a simple geometric 
expression for the RGS probability and discuss the effective impact 
parameters in exclusive diffraction. We then evaluate the RGS probability 
numerically, estimate the uncertainty of the numerical predictions
due to the phenomenological input, and emphasize the crucial role of the 
black--disk limit in stabilizing the numerical predictions.
We also comment on the results for the RGS obtained within 
the eikonalized Pomeron model for soft interactions \cite{Khoze:2000wk} 
from the perspective of our approach. In Section~\ref{sec:beyond} 
we discuss various effects beyond the approximation of independence 
hard and soft interactions in exclusive diffraction. 
We first point out that hard screening corrections may reduce 
the diffractive cross section beyond the RGS probability due to soft
interactions. We then discuss the effect of correlations between 
hard and soft interactions on the RGS probability, considering
two specific mechanisms --- diffractive scattering from the long--range 
pion field, and short--range transverse correlations between partons.
In Section~\ref{sec:differential} we work out the dependence of the 
exclusive diffractive cross section on the final proton transverse momenta.
We discuss which experimentally observable features of this dependence 
furnish useful tests of the diffractive reaction mechanism,
and how one can extract information about the gluon GPD.
In Section~\ref{sec:discussion} we summarize our results. 
We comment on the implications for the Higgs boson search, and
on the experimental feasibility of measuring the transverse momentum
dependence of exclusive diffraction with the planned forward 
detectors at the LHC.
\section{Black--disk limit in $pp$ elastic scattering}
\label{sec:transverse_soft}
Information on the transverse radius of strong interactions at 
high energies comes mostly from measurements of the $t$--dependence
of the differential cross section for $pp$ and $\bar p p$ elastic scattering. 
Combining these data with those on the $pp/\bar pp$ total cross section, 
and implementing theoretical constraints following from the unitarity 
of the $S$--matrix, one can reconstruct the complex $pp$ elastic scattering 
amplitude, $T_{\text{el}}(s, t)$; see \textit{e.g.}\ 
Refs.~\cite{Block:1998hu,Bourrely:2002wr,Islam:2002au}.
At high energies, $s \gg |t|$, angular momentum conservation in the CM 
frame implies that the scattering
amplitude is effectively diagonal in the impact parameter of the
colliding $pp$ system. It is convenient to represent it as a Fourier
integral over a transverse coordinate variable, $\bm{b}$,
\be
T_{\text{el}} (s, t = -\bm{\Delta}_\perp^2) 
&=& \frac{i s}{4\pi} \int d^2 b \;
e^{-i (\bm{\Delta}_\perp \bm{b})}
\; \Gamma (s, \bm{b}) ,
\label{Gamma_def}
\ee
where $\Gamma$ is the (dimensionless) profile function. 
One can then express the elastic, total, and inelastic
(total minus elastic) $pp$ cross sections in terms of the 
profile function as
\begin{equation}
\left. \begin{array}{l} 
\sigma_{\text{el}}(s) \\[1ex]
\sigma_{\text{tot}}(s) \\[1ex]
\sigma_{\text{inel}}(s)
\end{array}
\right\}
\;\; = \;\; \int d^2 b \; \times 
\left\{ \begin{array}{l} 
|\Gamma (s,\bm{b})|^2 , \\[1ex]
2 \, \text{Re} \, \Gamma (s,\bm{b}) , \\[1ex]
\left[ 1 - |1-\Gamma (s,\bm{b})|^2 \right] .
\end{array}
\right.
\label{unitarity}
\end{equation}
The functions on the R.H.S.\ describe the distribution of the respective
cross sections over $pp$ impact parameters, $b \equiv |\bm{b}|$
\footnote{For $pp$ scattering without transverse polarization effects, 
which we consider here,
the profile function depends only on the modulus of the transverse
coordinate, $b = |\bm{b}|$. We nevertheless regard it as a function 
of the vector variable, $\bm{b}$, to facilitate later usage when we 
consider convolutions in the transverse coordinate variable.}. 
In particular, we note that the combination
\begin{equation}
|1 - \Gamma (s, \bm{b})|^2
\label{P_noin}
\end{equation}
can be interpreted as the probability for ``no inelastic interaction''
in a $pp$ collision at impact parameter $b$; this combination plays
an important role in our calculation of the RGS probability (see
Section~\ref{sec:cross_section} below) \footnote{Notice 
that $1 - \Gamma(s, \bm{b})$ can be interpreted as the impact 
parameter representation of the $S$--matrix element, which is related 
to the scattering amplitude by Eqs.~(\ref{S_minus_1_me}) and 
(\ref{S_minus_1_t}).}.
A measure of the transverse size of the proton is the logarithmic 
$t$--slope of the elastic $pp$ cross section at $t = 0$,
\beq
B \;\; \equiv \;\; \frac{d}{dt} \left[ 
\frac{d\sigma_{\text{el}} /dt \, (t)}{d\sigma_{\text{el}} /dt \, (0)} 
\right]_{t = 0}.
\label{B_zero}
\eeq
At high energies, where the elastic amplitude is predominantly 
imaginary, and $\Gamma$ is real, $B$ is equal to 
half the average squared impact parameter in the total 
$pp$ cross section,
\beq
B \;\; \approx \;\; 
\frac{\langle b^2 \rangle_{\text{tot}}}{2} \; 
\;\; \equiv \;\; \frac{1}{2} \frac{\int d^2 b \, b^2 \, 
2 \, \text{Re} \, \Gamma (s,\bm{b})}
{\int d^2 b \; 2 \, \text{Re} \, \Gamma (s,\bm{b})} ,
\eeq
which may be associated with the transverse area of the 
individual protons. The data show that the slope increases 
with the CM energy as
\be
B(s) \;\; =\;\; B(s_0) \; + \; 2 \alpha^{\prime}\; \ln (s/s_0),
\label{alpha_prime}
\ee
where $\alpha^{\prime} \approx 0.25 \, \text{GeV}^{-2}$. 
In the Pomeron exchange parametrization of the $pp$ elastic amplitude
this constant is identified with the slope of the Pomeron trajectory.

In Gribov's parton picture of high--energy hadron--hadron 
interactions \cite{Gribov:1973jg}, the transverse size of the proton 
in $pp$ elastic scattering can be directly associated with the average 
transverse radius squared of the distribution of soft partons 
mediating the soft interactions,
\beq
B \;\; = \;\; \langle \rho^2 \rangle_{\text{soft}}.
\eeq
Here and in the following, we use $\rho \equiv |\bm{\rho}|$ to denote 
the transverse distance of partons from the center of the proton, 
and $b = |\bm{b}|$ for the impact parameter of the $pp$ collision. 
The growth of the proton's transverse size with energy is explained 
as the result of random transverse displacements in the successive 
decays generating the distribution of soft partons (Gribov diffusion).
Below we shall compare this distribution of soft partons to the 
distribution of hard partons probed in hard exclusive processes
(see Section~\ref{sec:transverse_hard}). 

Parametrizations of the available data indicate that at energies above 
the Tevatron energy, 
$\sqrt{s} \gtrsim \sqrt{s_{\text{Tevatron}}} = 2\, \text{TeV}$, 
the profile function at small impact parameters approaches
\beq
\Gamma (s, b) \;\; \rightarrow \;\; 1
\hspace{3em} \text{for} \;\; b < b_0 (s) .
\eeq
This corresponds to unit probability for inelastic scattering
for impact parameters $b < b_0 (s)$, \textit{cf.}\ Eqs.~(\ref{unitarity})
and (\ref{P_noin}), similar to the scattering of a pointlike object from a 
black disk of radius $b_0$, and is referred to as the black--disk limit (BDL) 
\cite{Frankfurt:2003td,Frankfurt:2004ti,Frankfurt:2005mc}.

The approach to the BDL in central $pp$ scattering at high energies 
is a general prediction of QCD, independent of detailed assumptions
about the dynamics. Studies of the interaction of small--size color
dipoles with hadrons, based on QCD factorization in the leading
$\log Q^2$ approximation, show that the BDL is attained at high energies 
as a result of the growth of the gluon
density at small $x$ due to DGLAP evolution \cite{Frankfurt:2003td}. 
This result can be used to estimate the interaction of 
leading projectile partons with the small--$x$ gluons in the target
in $pp$ scattering; one finds that there is no chance for the
projectile wavefunction to remain coherent in small impact parameter
scattering at TeV energies \cite{Frankfurt:2004ti,Frankfurt:2005mc}. 
Similar reasoning allows one to predict the growth of the size of the 
black region, $b_0$, with $s$ \cite{Frankfurt:2004ti,Frankfurt:2005mc}.
As a by--product, these arguments explain why the observed 
coefficient in the Froissart formula for the total cross sections is
significantly smaller than that derived from the general principles 
of analyticity of the amplitude in momentum transfer and unitarity of the
$S$--matrix \cite{Martin:1962rt}. We note that the need for the approach 
to the BDL in high--energy scattering at central impact parameters was 
understood already in the pre--QCD period within the Pomeron calculus,
where it was noted that this phenomenon resolves the apparent contradiction 
between the formulae of the triple--Pomeron limit and the unitarity of the
$S$--matrix, especially in models where the Pomeron intercept, 
$\alpha_P(0)$, exceeds unity \cite{Marchesini:1976hw}.

%
%
\begin{figure}
\includegraphics[width=8cm]{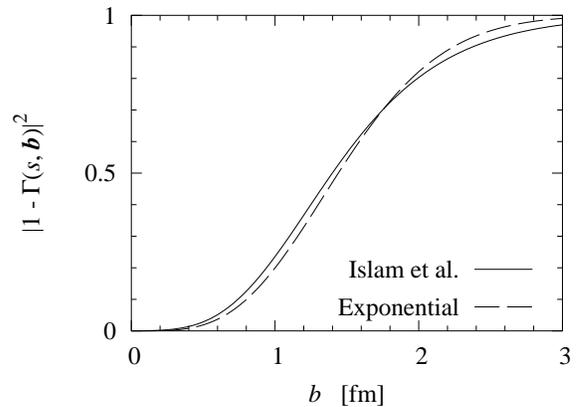}
\caption[]{The probability distribution for no inelastic 
interaction, $|1 - \Gamma (s, \bm{b})|^2$, Eq.~(\ref{P_noin}),
as a function of $b \equiv |\bm{b}|$, 
at the LHC energy ($\sqrt{s} = 14 \, \text{TeV}$),
as computed with different parametrizations of the 
$pp$ elastic scattering amplitude. Solid line: Parametrization of 
Islam \textit{et al.}\ \cite{Islam:2002au} (``diffractive part'' only). 
Dashed line: Exponential parametrization, Eq.~(\ref{Gamma_gaussian}),
with $\Gamma_0 = 1$ (BDL), \textit{cf.}\ Eq.~(\ref{Gamma_0_BDL}), and 
$B = 21.8 \, \text{GeV}^{-2}$.}
\label{fig:gamma}
\end{figure}
For our studies of diffractive $pp$ scattering it will be useful 
to have a simple analytic parametrization of the $pp$ elastic amplitude 
at the LHC energy, which incorporates the approach to the BDL at small 
impact parameters. The $t$--dependence of the $pp$ elastic scattering 
cross section for $|t| \lesssim 1 \, \text{GeV}^2$ over the measured
energy range is reasonably described by an exponential shape,
\beq
\frac{d\sigma_{\text{el}}}{dt} \;\; \propto \;\; \exp \left[ B(s) \, t 
\right] ,
\eeq
where $B(s)$ represents an effective slope, to be distinguished from the
``exact'' slope at $t = 0$, Eq.~(\ref{B_zero}). A parametrization of the 
$pp$ elastic amplitude which reproduces this dependence is
\beq
T_{\text{el}} (s, t) \;\; = \;\; \frac{is}{8\pi} \; \sigma_{\text{tot}}(s)
\; \exp \left[ \frac{B(s) \, t}{2} \right] ,
\label{A_exponential}
\eeq
corresponding to
\be
\Gamma (s, \bm{b}) \;\; = \;\; 
\Gamma_0 (s) \; \exp \left[ -\frac{\bm{b}^2}{2 B(s)} \right]
\label{Gamma_gaussian}
\ee
with
\beq
\Gamma_0 (s) \;\; \equiv \;\; \Gamma (s, \bm{b} = 0)
\;\; = \;\; \frac{\sigma_{\text{tot}}(s)}{4\pi B(s)} .
\eeq
Equation~(\ref{A_exponential}) takes into account that the
amplitude at high energies is predominantly imaginary, and satisfies
the optical theorem for the total cross section,
$\sigma_{\text{tot}}(s) = (8\pi /s) \, \text{Im}\, T_{\text{el}} (s, t = 0)$.
We may now incorporate the constraint of the BDL at small impact parameters 
by replacing
\beq
\Gamma_0 \;\; \rightarrow  1.
\label{Gamma_0_BDL}
\eeq
The value of $B$ we determine by comparing the profile function
(\ref{Gamma_gaussian}) with phenomenological parametrizations
of the data, extrapolated to the LHC energy, which gives
\beq
B \;\; \approx \;\; 
20 \, \text{GeV}^{-2} \hspace{3em} (\sqrt{s} = 14 \, \text{TeV}) .
\label{B_LHC}
\eeq
In particular, with $B = 21.8 \, \text{GeV}^{-2}$ we obtain excellent
agreement with the Regge parametrization of Ref.~\cite{Khoze:2000wk}.
Figure~\ref{fig:gamma} shows the probability for no inelastic interaction,
$|1 - \Gamma (s, \bm{b})|^2$, Eq.~(\ref{P_noin}), computed with the
phenomenological parametrization of Ref.~\cite{Islam:2002au} and 
our exponential parametrization incorporating the BDL, 
Eqs.~(\ref{Gamma_gaussian}) and (\ref{Gamma_0_BDL}).
One sees that the simple exponential parametrization is a
reasonable overall approximation to the phenomenological parametrization 
over the $b$--range shown in Fig.~\ref{fig:gamma}.
\section{Transverse spatial distribution of gluons}
\label{sec:transverse_hard}
Information about the transverse structure of hard interactions
comes from studies of hard exclusive processes in $ep$ scattering,
such as meson electroproduction or virtual Compton scattering.
Such processes probe the GPDs in the proton, whose Fourier transform 
with respect to the transverse momentum transfer to the proton 
describes the spatial distribution of quarks and gluons
in the transverse plane; see Refs.~\cite{Diehl:2003ny,Belitsky:2005qn}
for a review. In this section we summarize what is
known about the gluon GPD at small $x$ from theoretical
considerations, and from measurements of $J/\psi$ photoproduction 
and other processes at HERA and in fixed--target experiments.

The gluon GPD can be formally defined as the transition matrix 
element of the twist--2 QCD gluon operator between proton states of 
different momenta, $p$ and $p'$. Physically, it describes the amplitude 
for a fast--moving proton to ``emit'' and ``absorb'' a gluon with given 
longitudinal momenta, with transverse momenta (virtualities) 
integrated over up to some hard scale, $Q^2$, and a certain invariant 
momentum transfer to the proton, $t \equiv (p' - p)^2$. 
The choice of longitudinal momentum
variables is a matter of convention. Instead of the initial and final
gluon momentum fractions (with respect to the initial proton momentum),
$x$ and $x'$, we use as independent variables the initial gluon 
momentum fraction, $x$, and the fractional longitudinal momentum transfer 
to the proton (``skewness'') \footnote{In most of the recent literature the
skewness is defined such that $2\xi = x' - x$. In the present context 
it is convenient to omit the factor 2 and define the skewness
as in Eq.~(\ref{xi_def}).}
\beq
\xi \;\; \equiv \;\; x - x',
\label{xi_def}
\eeq
and denote the gluon GPD by
\beq
H_g (x, \xi, t; Q^2) .
\eeq
In the limit of zero momentum transfer, the gluon GPD reduces to the 
usual gluon momentum density in the proton \footnote{In the present context 
it is convenient to define the gluon GPD as corresponding to the gluon 
momentum density, $x G(x)$, not the number density, $G(x)$, in the limit 
of zero momentum transfer.}
\beq
H_g (x, \xi = 0, t = 0; Q^2) \;\; = \;\; x G(x, Q^2 ) .
\label{H_g_forward}
\eeq
For discussing the $t$--dependence of the gluon GPD 
it is convenient to write it in the form
\beq
H_g (x, \xi, t; Q^2) \;\; = \;\;  H_g (x, \xi, t = 0; Q^2) \; 
F_g (x, \xi, t; Q^2),
\label{twogl_def}
\eeq
where the function $F_g$ is known as the ``two--gluon formfactor'' of 
the proton and satisfies $F_g (t = 0) = 1$. Note that the two--gluon
formfactor still depends on $x$ and $\xi$, \textit{i.e.}, 
Eq.~(\ref{twogl_def}) does not imply naive factorization of the 
$t$--dependence from that on the partonic variables.

The dependence of the gluon GPD on the QCD scale, $Q^2$, is governed 
by the QCD evolution equations. In applications to production of 
fixed--mass systems in high--energy $ep$ or $pp$ collisions with 
$M^2 \ll s$ (such as Higgs boson production at the LHC) we shall be
interested in the gluon GPD in the region where
\beq
x, x' \;\; \ll \;\; 1 ,
\eeq
while at the same time $Q^2$ is much larger than the typical hadronic 
mass scale, $Q^2 \gg 1\, \text{GeV}^2$. In this region the gluon GPD
can be calculated by applying QCD evolution to a ``primordial'' 
distribution at a low scale, $Q_0^2$, in which one neglects the
skewness, $\xi = 0$, or $x = x'$ (diagonal approximation)
\cite{Frankfurt:1997ha,Shuvaev:1999ce}. QCD evolution degrades 
the individual gluon momentum fractions with increasing $Q^2$, 
while their difference, $\xi = x - x'$, remains 
fixed by kinematics, being equal to the longitudinal momentum
transfer to the proton; as a result, the primordial GPD at the 
low scale is sampled in the region $|x - x'| \ll x, x'$, 
where the diagonal approximation is justified. In the diagonal
approximation, the GPD at $t = 0$ is completely determined
by the usual gluon density, \textit{cf.}\ Eq.~(\ref{H_g_forward}),
leaving only the $t$--dependence (two--gluon formfactor) and its
correlation with $x$ up to modeling. This approximation
makes for a great simplification in applying GPDs at small $x$,
and has been used extensively in the analysis of exclusive 
electroproduction processes; see Ref.~\cite{Frankfurt:2005mc}
for a review.

For $x \ll 1$, the two--gluon formfactor permits a
simple interpretation in terms of a spatial distribution of gluons
in the proton. For $\xi \ll 1$, the invariant momentum transfer
is dominated by the transverse momentum transfer between the
proton states,
\beq
t \;\; \approx \;\; -\bm{\Delta}_\perp^2,
\hspace{3em} \bm{\Delta}_\perp \;\; \equiv \;\; 
\bm{p}'_\perp - \bm{p}_\perp .
\eeq
The two--gluon formfactor can be represented as a 
Fourier integral over a transverse coordinate variable, $\bm{\rho}$,
\be
F_g (x, \xi, t = -\bm{\Delta}_\perp^2; Q^2) &=& 
\int d^2 \rho \; e^{-i (\bm{\Delta}_\perp \bm{\rho})} 
\nonumber \\
&\times& F_g (x, \xi, \bm{\rho}; Q^2) .
\label{rhoprof_def}
\ee
For economy of notation we use the same symbol for the
two--gluon formfactor and its Fourier transform, distinguishing
the two functions by their arguments. The $\bm{\rho}$--dependent
function describes the spatial distribution of gluons 
in the proton in the transverse plane; see Ref.~\cite{Burkardt:2002hr}
for a review. For $\xi = 0$ (\textit{i.e.}, $x' = x$) it is positive 
definite and can be interpreted probabilistically as the gluon density 
at transverse position $\bm{\rho}$ \cite{Pobylitsa:2002iu}; 
for $\xi \neq 0$ it describes the non-diagonal transition matrix element 
of the gluon density \cite{Diehl:2002he}. A measure of the gluonic transverse
size of the nucleon for given $x$ and $Q^2$ is the average of $\rho^2$,
which is proportional to the $t$--slope of the two--gluon formfactor
at $t = 0$,
\be
\langle \rho^2 \rangle_g \, (x, Q^2) &\equiv&
\int d^2 \rho \; \rho^2 \; F_g (x, \, \xi = 0, \, \bm{\rho}; \, Q^2 )
\\
&=&
4 \, \frac{\partial F_g}{\partial t} \,  (x, \, \xi = 0, \, t = 0; \, Q^2).
\ee

The two--gluon formfactor of the nucleon, and hence the transverse
spatial distribution of gluons, can directly be extracted from the
$t$--dependence of the differential cross section for hard exclusive
vector meson production processes probing the gluon GPD. 
QCD factorization implies that the $t$--dependence of the cross 
section resides solely in the gluon GPD,
\beq
\left( \frac{d\sigma}{dt} \right)^{\gamma^* p \rightarrow V p}
\;\; \propto \;\; F_g^2 (x, \xi, t; Q^2) ,
\eeq
up to small higher--twist corrections due to the finite size of the 
produced vector meson \cite{Frankfurt:2005mc}. In particular, the 
$t$--slope at $t = 0$ is proportional to the proton's average 
gluonic transverse size, 
\beq
B_g \;\; \equiv \;\; \frac{d}{dt} \left[ 
\frac{d\sigma/dt \, (t)}{d\sigma/dt \, (0)} 
\right]^{\gamma^* p \rightarrow V p}_{t = 0}
\;\; = \;\; \frac{\langle \rho^2 \rangle_g}{2} .
\eeq
A crucial test of the 
applicability of QCD factorization is provided by the observed
convergence of the $t$--slopes of various gluon--dominated
vector meson production processes ($J/\psi, \rho, \phi$) at large
$Q^2$; see Ref.~\cite{Frankfurt:2005mc} for a detailed discussion.

A particularly clean probe of the two--gluon formfactor is
$J/\psi$ photoproduction, the $t$--dependence of which has been
measured in experiments at the HERA collider 
($x \sim 10^{-2} - 10^{-4}$) \cite{Aktas:2005xu,Chekanov:2004mw},
the FNAL fixed--target experiment 
($\langle x \rangle \sim 5 \times 10^{-2}$) \cite{Binkley:1981kv},
and a number of fixed--target experiments at lower energies
($x \sim 10^{-1}$); see Refs.~\cite{Frankfurt:2002ka,Strikman:2004km} 
for a review of the data. This process probes the two--gluon formfactor 
at an effective scale $Q^2 \approx 3 \, \text{GeV}^2$. Analysis of the data, 
combined with theoretical investigations, has produced a rather detailed 
picture of the gluonic transverse size of the nucleon and its 
$x$--dependence \cite{Frankfurt:2005mc}. For $x \sim 0.1 - 0.3$, 
the gluonic transverse size suggested by the fixed--target data is 
$\langle \rho^2 \rangle_g \approx 0.25 \, \text{fm}^2$, close 
to $2/3$ times the proton's axial charge radius, $\langle r^2 \rangle_A$. 
Between $x \sim 10^{-1}$ and $x \sim 10^{-2}$, $\langle \rho^2 \rangle_g$ 
increases by $\sim 30\%$. This can be explained by the contribution of the 
nucleon's pion cloud to the gluon density at large transverse distances,
$\rho \sim 1/(2 M_\pi )$, which is dynamically suppressed for 
$x > M_\pi / M_N$ and reaches its full strength for 
$x \ll M_\pi / M_N$ \cite{Strikman:2003gz}. Finally, over 
the HERA range, $x \sim 10^{-2} - 10^{-4}$, the gluonic 
transverse size exhibits a logarithmic growth with $1/x$,
\beq
\langle \rho^2 \rangle_g \;\; = \;\; 
\langle \rho^2 \rangle_g (x_0) \; + \; 4 \, \alpha'_g \, \ln\frac{x_0}{x} 
\hspace{2em} (x < x_0 \approx 10^{-2}),
\label{alpha_g}
\eeq
with a rate, $\alpha'_g$, considerably smaller than that governing the 
growth of the proton's transverse size in $pp$ elastic scattering,
which is dominated by soft interactions,
\beq
\alpha'_g \;\; \ll \;\; \alpha' .
\label{alpha_prime_comparison}
\eeq
A recent analysis of the H1 data finds $\alpha'_g = 0.164 \pm 0.028 
(\mbox{stat}) \pm 0.030 (\mbox{syst}) \, \text{GeV}^{-2}$ for 
$J/\psi$ photoproduction
and $0.019 \pm 0.139 (\mbox{stat}) \pm 0.076 (\mbox{syst}) \,
\text{GeV}^{-2}$ 
for electroproduction \cite{Aktas:2005xu}; 
an analysis of ZEUS electroproduction data
quotes $\alpha'_g = 0.07 \pm 0.05 (\mbox{stat}) { }^{+0.03}_{-0.04}
(\mbox{syst}) \, \text{GeV}^{-2}$ \cite{Chekanov:2004mw}, significantly
smaller than the soft value $\alpha' = 0.25 \, \text{GeV}^{-2}$.
The smaller rate of growth of the nucleon's size in hard interactions 
can qualitatively be explained by the suppression of Gribov diffusion 
in the decay of hard (highly virtual) partons as compared to soft partons.

%
%
\begin{figure}
\includegraphics[width=7cm]{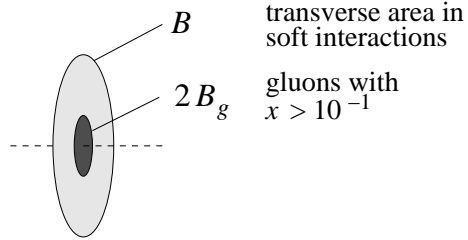}
\caption[]{The ``two--scale picture'' of the transverse 
structure of the proton in high--energy collisions.}
\label{fig:twoscale}
\end{figure}
A crucial observation is that the transverse area occupied by
partons with $x \gtrsim 10^{-1}$ is much smaller than the
transverse area associated with the proton in soft interactions
(see Fig.~\ref{fig:twoscale}),
\beq
\langle \rho^2 \rangle_g \, (x \gtrsim 10^{-1})
\;\; \ll \;\; \langle \rho^2 \rangle_{\text{soft}},
\eeq
or
\beq
2 B_g \;\; \ll \;\; B .
\eeq
In high--energy $pp$ collisions with hard partonic processes one 
is thus dealing with a two--scale picture of the transverse structure 
of the proton. Moreover, when considering the production of a heavy
particle with fixed mass, $m_H$, in a partonic process with 
$x_{1, 2} \sim m_H / \sqrt{s}$, the ``soft'' area
of the proton increases with $s$ faster than the ``hard'' area
(which changes as a result of the decrease of $x$), 
because $\alpha' > \alpha'_g$, \textit{cf.}\ 
Eq.~(\ref{alpha_prime_comparison}). Thus, the difference of the
two areas becomes even more pronounced with increasing energy.

%
%
\begin{figure}
\includegraphics[width=8cm]{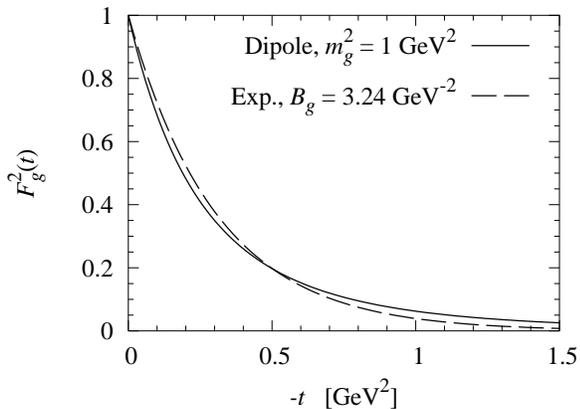}
\caption[]{Comparison of the dipole (solid line) and exponential
(dashed line) parametrizations of the two--gluon formfactor, with 
the parameters related by Eq.~(\ref{dip_exp}). Shown is the 
squared two--gluon formfactor, $F_g^2 (t)$, for both parametrizations,
as corresponds to the $t$--dependence of the cross section for 
$J/\psi$ photoproduction (details see text).}
\label{fig_fg_dip_exp}
\end{figure}
For our studies of hard processes in diffractive $pp$ scattering
we require a parametrization of the $t$--dependence of the
two--gluon formfactor, \textit{viz.}\ the shape of the transverse 
spatial distribution of gluons. The $x$--values probed in 
Higgs production at central rapidities are $x \sim 10^{-2}$
at the LHC energy. Taking into account the effect of DGLAP evolution, 
even larger values of $x$ are probed when parametrizing the two--gluon
formfactor at the $J/\psi$ production scale, $Q^2 \sim 3 \, \text{GeV}^2$
(for a general discussion of the effect of DGLAP evolution on the transverse
spatial distribution of gluons, see Ref.~\cite{Frankfurt:2003td}). 
We thus need to look at the $J/\psi$ photoproduction data at 
$x \gtrsim 10^{-2}$, which are probed in fixed--target experiments.

Theoretical arguments suggest that the two--gluon formfactor at 
$x \gtrsim 10^{-1}$ should be close to the axial formfactor, 
which is well described by a dipole form (we omit all arguments
except $t$),
\beq
F_g (t) \;\; = \;\; \frac{1}{(1 - t/m_g^2)^2} , 
\label{twogl_dipole}
\eeq
with $m_g^2 \approx 1 \, \text{GeV}^2$ \cite{Frankfurt:2002ka}. 
The corresponding transverse spatial distribution of gluons is given by
\beq
F_g (\bm{\rho}) \;\; = \;\; \frac{m_{g}^2}{2\pi}
\; \left(\frac{m_{g} \rho}{2}\right) \; K_1 (m_{g} \rho ) .
\label{spatial_dipole}
\eeq
We also consider an exponential parametrization
of the two--gluon formfactor,
\beq
F_g (t) \;\; = \;\; \exp (B_g t/2) , 
\label{twogl_exponential} 
\eeq
corresponding to
\beq
F_g (\bm{\rho} ) \;\; = \;\; 
\frac{\exp \left[ -\bm{\rho}^2 / (2 B_g)\right]}
{2 \pi B_g} .
\label{spatial_gaussian} 
\eeq
The relation between the parameters of the dipole and exponential
parametrization which would follow from identifying
$\langle \rho^2 \rangle = 4 \, dF_g / d t (t = 0)$
is $B_g = 4/m_g^2$. Better overall agreement between the squared 
formfactors for $|t| < 1\, \text{GeV}^2$ is obtained for somewhat 
smaller values of $B_g$. Matching the squared formfactors at 
$|t| = 0.5\, \text{GeV}^2$ we obtain 
\beq
B_g \;\; = \;\; \frac{3.24}{m_g^2} ,
\label{dip_exp}
\eeq
see Figure~\ref{fig_fg_dip_exp}. It was shown in Ref.~\cite{Strikman:2004km} 
that both the dipole with $m_g^2 = 1.1\,\text{GeV}^2$ and the
exponential with $B_g = 3.0 \, \text{GeV}^2$ given by Eq.~(\ref{dip_exp}) 
describe well the $t$--dependence of the data from the FNAL E401/E458 
experiment at $\langle E_\gamma \rangle = 100\; {\rm GeV}$ in which the 
recoiling proton was detected \cite{Binkley:1981kv}. We also note
that this value of $B_g$ is consistent with what
one obtains from the extrapolation of the HERA data towards larger
$x$, using Eq.~(\ref{alpha_g}) with the measured $\alpha'_g$.
We shall use the dipole, Eqs.~(\ref{twogl_dipole}), with 
$m_g^2 = 1 \, \text{GeV}^2$ and the exponential, 
Eq.~(\ref{twogl_exponential}), with $B_g = 3.24 \, \text{GeV}^{-2}$,
as our standard parametrizations for calculations in the kinematics of
Higgs production at the LHC below; comparison between the two will
allow us to estimate the uncertainty of our numerical predictions with
respect to the shape of the two--gluon formfactor.
\section{Theory of rapidity gap survival}
\label{sec:gap_survival}
We now outline the basic steps in the calculation of the 
amplitude of double--gap exclusive diffractive processes 
(\ref{exclusive_diffraction}),
and develop the physical picture of RGS.
The underlying idea of our approach is that hard and soft interactions 
are approximately independent, because they happen over widely different 
distance-- and time scales.
\subsection{Hard scattering process}
\begin{figure}
\includegraphics[width=5cm]{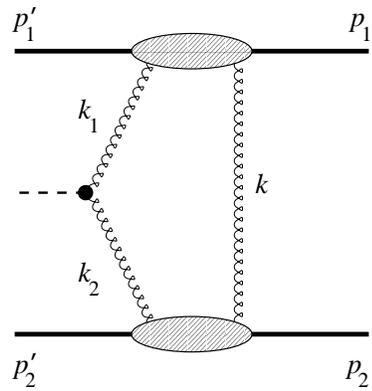}
\caption[]{The hard scattering process in double--gap exclusive diffractive 
Higgs boson production (\ref{exclusive_diffraction}). Two gluons
are exchanged between the protons. The gluon--Higgs coupling is
indicated as a local vertex. The upper and lower blobs
denote the gluon--proton scattering amplitude, which can be 
calculated in terms of the gluon GPD in the proton.}
\label{fig:hard}
\end{figure}

In the first step, one calculates the amplitude for double--gap
diffractive production of the high--mass system due to hard interactions. 
For definiteness, we shall refer in the following to Higgs boson production, 
keeping in mind that the discussed mechanism applies to production 
of other high--mass states as well (dijets, heavy quarkonia, \textit{etc.}).
According to electroweak theory, the Higgs boson is produced 
predominantly through its coupling to gluons via a quark loop;
for a review and references see Ref.~\cite{Gunion:1992hs}. 
In contrast to inclusive production, 
the amplitude for double--gap diffractive production is 
in the lowest order in the QCD running coupling constant, $\alpha_s$, 
given by the exchange of two gluons with vacuum quantum numbers 
in the $t$--channel (see Fig.~\ref{fig:hard}). The Higgs boson 
is radiated from one of the gluon lines. The role of the second 
exchanged gluon is to neutralize the color charge in order to avoid gluon 
bremsstrahlung. However, global color neutrality alone is not sufficient. 
To suppress radiation, one must require that color be screened 
locally in space--time. Conversely, this means that the selection
of a diffractive process, without accompanying radiation, guarantees
some degree of localization of the exchanged system.

Operationally, the localization of the exchanged two--gluon system is
ensured by Sudakov formfactors, which suppress configurations with
low virtualities prone to emit gluon bremsstrahlung. The actual 
calculation of the hard scattering amplitude including Sudakov 
suppression is a challenging problem, which was addressed in
various approximations in Refs.~\cite{Khoze:1997dr,Khoze:2000cy}. 
Fortunately, for our purposes we do not need to solve this problem 
at a fully quantitative level, as only a few qualitative aspects of 
the hard scattering process turn out to be essential for the 
physics of RGS.

To discuss the hard scattering process, it is natural to perform 
a Sudakov decomposition of the four--momenta, using the
initial proton momenta, $p_1$ and $p_2$, as basis vectors, with 
$2 (p_1 p_2) = s$ (we neglect the proton mass). As the transverse
momenta of the final--state protons are small compared to the 
Higgs mass, we can expand the final proton four--momenta as
\be
p_1' &=& (1 - \xi_1) p_1 \; + \; p_{1\perp}' ,
\nonumber \\
p_2' &=& (1 - \xi_2) p_2 \; + \; p_{2\perp}' ,
\ee
where $(p_{1\perp}', p_1) = (p_{1\perp}', p_2) = 0$ \textit{etc.}, 
and $\xi_{1, 2}$ parametrize the longitudinal momentum loss
[\textit{cf.}\ Eq.~(\ref{xi_def}) and the footnote before it],
\beq
\xi_{1, 2} \;\; = \;\; \frac{m_H}{\sqrt{s}} \, e^{\pm y},
\label{xi_kinematics}
\eeq
where $y$ is the rapidity of the produced Higgs boson.
Assuming a Higgs mass of the order of $100 - 200 \, \text{GeV}$, 
the typical values of $\xi_{1, 2}$ are of the order of $10^{-2}$
for production at central rapidities at the LHC 
($\sqrt{s} = 14 \, \text{TeV}$). 

%
%
Consider now the two--gluon exchange process of Fig.~\ref{fig:hard}
as a Feynman diagram, in which the upper and lower blobs denote the
gluon--proton scattering amplitudes, to be specified in more detail 
below. The four--momenta of the gluons 
coupling to the Higgs we parametrize as
\be
k_1 &=& x_1 p_1 \; + \; x_2' p_2 \; + \; k_\perp - p_{1\perp}', 
\label{k_1_Sudakov}
\\
k_2 &=&  x_1' p_1 \; + \; x_2 p_2 
\; - \; k_\perp - p_{2\perp}' .
\label{k_2_Sudakov}
\ee
The four--momentum of the screening gluon then follows from
four--momentum conservation, 
\beq
k \;\; = \;\; x_1' p_1 \; + \; x_2' p_2 \; + \; k_\perp ,
\eeq
where
\be
x_1' &\equiv& \xi_1 - x_1 , 
\\
x_2' &\equiv& \xi_2 + x_2 . 
\ee
We want to identify the dominant region of integration in the 
loop integral. First, analogy with inclusive production of heavy particles
at central rapidities suggests that the momentum fractions of the 
gluons producing the Higgs boson (with respect to their parent protons) 
are practically the same as given by the naive parton model estimate, 
in which one neglects the transverse momenta and virtualities 
of the annihilating gluons,
\beq
x_{1,2} \;\; \sim \;\; \frac{m_H}{\sqrt{s}} \, e^{\pm y} .
\label{x_12}
\eeq
That is, the momentum fractions of the annihilating gluons are equal
to the protons' fractional longitudinal momentum loss, 
$x_{1, 2} \approx \xi_{1, 2}$.  Second, in the case of double--gap
diffractive production the Sudakov formfactor associated with the
Higgs boson vertex, which accounts for the absence of gluon
bremsstrahlung, restricts the (spacelike) virtualities of the
annihilating gluons and their transverse momenta to values of the
order of some ``intermediate'' hard scale,
\beq
-k_{1, 2}^2, \; -k_\perp^2 \;\; \sim \;\; Q_{\text{int}}^2,
\label{k2_Q2}
\eeq
with
\beq
m_H^2/4  \;\; \gg \;\; Q_{\text{int}}^2 \;\; \gg \;\; \Lambda_{\text{QCD}}^2 .
\label{Q2_def}
\eeq
This was demonstrated explicitly in Ref.~\cite{Khoze:1997dr}, where
the distribution of transverse momenta in the loop integral was studied
in a model which included the LO Sudakov formfactor associated 
with the $ggH$ vertex; see also Ref.~\cite{Khoze:2000cy} 
\footnote{The Sudakov formfactor associated 
with the $ggH$ vertex accounts for the absence of gluon bremsstrahlung 
with transverse momenta $|k_\perp| < |k_{\perp, \text{rad}}| < m_H/2$,
radiated from the annihilating gluons.
The requirement of absence of radiation from the screening gluon 
results  in an additional shift of the $k_\perp$ distribution towards 
larger values. However the amplitude for radiation from this line  
does not contain factors $\alpha_s \ln (k_\perp^2 /m_H^2)$, and is 
beyond the accuracy of the calculation of Ref.~\cite{Khoze:1997dr}.}.
Expressing now the virtualities of the annihilating gluons, $k_{1, 2}^2$,
in terms of the decompositions (\ref{k_1_Sudakov}) and
(\ref{k_2_Sudakov}), neglecting the proton transverse momenta
relative to $k_\perp$, we find that Eq.~(\ref{k2_Q2}) implies
\beq
x_{1, 2}' \;\; = \;\; \frac{k_{2, 1}^2 - k_\perp^2}{x_{2, 1} s}
\;\; \sim \;\; \frac{Q_{\text{int}}^2}{x_{2, 1} s}
\;\; \sim \;\; \frac{Q_{\text{int}}^2}{m_H \sqrt{s}}
\;\; \ll \;\; x_{1, 2} ,
\label{x_12_prime}
\eeq
\textit{i.e.}, the energy and longitudinal momentum fraction
of the screening gluon are substantially smaller than those of
the annihilating gluons. The screening gluon does not ``belong to''
any of the two protons; its momentum is predominantly transverse,
and it has spacelike virtuality,
\beq
-k^2 \;\; \approx \;\; -k_\perp^2 \;\; \sim \;\; Q_{\text{int}}^2 .
\eeq
In the annihilating gluons, on the other hand, longitudinal and
transverse momenta contribute in equal amounts to the virtuality,
\beq
-k_{1,2}^2 \;\; \approx \;\; x_{1, 2} x_{2, 1}' s - k_\perp^2 
\;\; \sim \;\; Q_{\text{int}}^2 .
\eeq
To summarize, the hard scattering process takes the form of the
exchange of two gluons with comparable virtualities $\sim Q_{\text{int}}^2$,
and transverse momenta $\sim Q_{\text{int}}$, between the protons. 
Of the two gluons, one carries substantial longitudinal momentum fraction 
of the proton, $\sim m_H/\sqrt{s}$, and annihilates with the
corresponding other to make the Higgs, the other gluon represents
a ``Coulomb--like'' exchange with small momentum fraction 
$\sim Q_{\text{int}}^2/(m_H \sqrt{s})$.

The important point about the two--gluon exchange process
is the appearance of the intermediate hard scale, $Q_{\text{int}}^2$, 
governing the virtualities and transverse momenta of the exchanged gluons.
This allows us to make a crucial simplification in the description
of the gluon--proton scattering amplitudes, which we have not yet
specified so far. Namely, we argue that, in a partonic description
of the proton, the gluon--proton scattering amplitude is dominated 
by the two gluons coupling to the same parton \cite{Brodsky:1994kf}. 
This approximation is analogous to the assumption of dominance of the 
``handbag graph'' in virtual Compton scattering at large photon virtuality, 
$Q^2$, which is well established and forms the basis of QCD factorization
for this process \cite{Mueller:1998fv,Abramowicz:1995hb,%
Ji:1996nm,Radyushkin:1997ki,Collins:1998be}.
In this approximation, the gluon--proton scattering amplitude, 
which is predominantly imaginary at high energies, can be calculated 
in terms of the generalized parton distributions (GPDs) --- here, 
predominantly, gluon distributions --- in the protons. 

We do not attempt to calculate the absolute normalization of the
amplitude for double--gap hard diffractive 
production through two--gluon exchange in terms of the GPDs; doing so
would require a substantially more accurate evaluation of the
two--gluon exchange graph than the qualitative estimates presented
above. Fortunately, for the theory of RGS, the only
information we require (in addition to the qualitative properties of
the hard process derived above) is the dependence of the double--gap
hard diffractive amplitude on the transverse momentum transfers to the
protons, $p_{1\perp}$ and $p_{2\perp}$. For sufficiently large scales
$Q_{\text{int}}^2$, this dependence is described by the GPDs, 
even if the gluon momentum fractions in the hard
amplitude and the virtuality of the exchanges are subject to
integration over a certain range, and determined only in
order--of--magnitude. Thus, we can state that the 
$p_{1\perp}$-- and $p_{2\perp}$ dependence of the double--gap 
hard diffractive amplitude is proportional to
\beq
T_{\text{hard}} \;\; \propto \;\; 
H_g (x_1, \xi_1, t_1; Q^2) \; H_g (x_2, \xi_2, t_2; Q^2) ,
\label{A_hard_GPD}
\eeq
where 
\beq
t_1 \; = \; p_{1\perp}^{\prime 2} \; < 0,
\hspace{2em}
t_{2} \; = \; p_{2\perp}^{\prime 2} \; < 0
\eeq
are the invariant momentum transfers to the proton. Here the longitudinal 
momentum transfers, $\xi_{1, 2}$, are kinematically fixed by
Eq.~(\ref{xi_kinematics}), while parton momentum fractions, 
$x_{1,2}$, are determined by Eq.~(\ref{x_12}) with accuracy
given by Eq.~(\ref{x_12_prime}). The resolution scale, $Q^2$, 
at which the GPD needs to be taken here, is parametrically of 
the order $Q_{\text{int}}^2$ [\textit{cf.}~Eq.~(\ref{Q2_def})], 
but numerically substantially larger,
\beq
Q^2 \;\; \sim \;\; \text{several times} \;\;\; Q_{\text{int}}^2 .
\eeq
This follows from the fact that, by convention, $Q^2$ determines
the upper limit of the transverse momenta in the parton
distribution, while $Q_{\text{int}}^2$ is a measure of the 
dominant (average) values in the distribution, which for a 
$1/k_\perp^2$ distribution is significantly lower than the upper limit.

%
%
\begin{figure*}
\includegraphics[width=13cm]{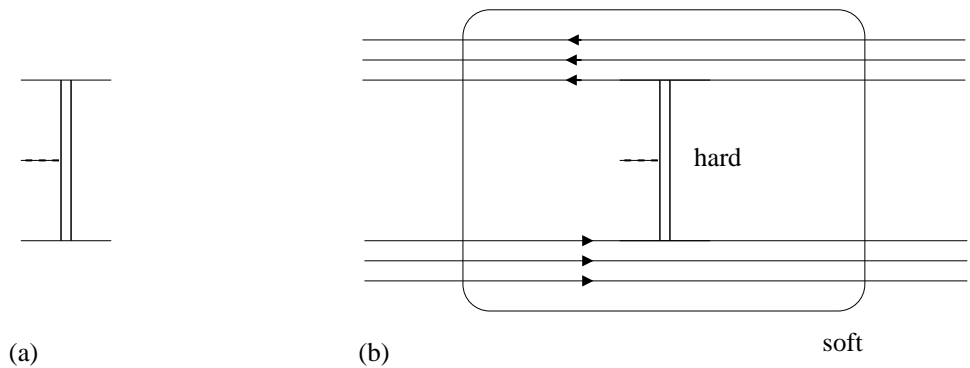}
\caption[]{Schematic illustration of hard and soft interactions
in the parton picture of double--gap exclusive diffractive $pp$ scattering. 
(a) The hard scattering process producing the large--mass system
(Higgs, dijet) is represented by a local operator in parton degrees 
of freedom. (b) Hard and soft interactions are approximately
independent since they happen over widely different time-- and
distance scales.}
\label{fig:hardsoft}
\end{figure*}
To conclude this discussion, some comments concerning QCD evolution
and the modeling of the GPDs are in order. First, in Higgs boson production
at the LHC we are dealing with the gluon GDP at $x, \xi \sim 10^{-2}$
and $x' \ll x$, where it can legitimately be calculated by applying QCD 
evolution to a diagonal GPD at a lower scale (\textit{cf.}\ the discussion 
in Section~\ref{sec:transverse_hard}). 
Second, at LHC energies the typical momentum fraction of the 
screening gluon reaches values $x_{1,2}' \sim 10^{-6}$
for $Q^2 \sim \text{few GeV}^2$, \textit{cf.}\ Eq.~(\ref{x_12_prime}).
For such low values of $x$ the use of DGLAP evolution in principle 
requires justification. However, as explained in detail in 
Ref.~\cite{Frankfurt:2005mc}, for such values of $x$ the kinematic 
conditions still restrict the actual number of radiated gluons to a few,
so that NLO DGLAP and resummed BFKL evolution give similar results,
see Ref.~\cite{Salam:2005yp} for a review. In this sense, the use of 
GPDs generated by DGLAP evolution seems to be appropriate.
Third, one may wonder about the breakdown of the pQCD calculation
of the hard reaction amplitude due to the growth of the gluon density 
at small $x$ within the DGLAP approximation --- the BDL, 
see Ref.~\cite{Frankfurt:2005mc} and references therein. 
This should not be a problem in the present context,
since the BDL affects only collisions at central impact
parameters, where the diffractive amplitude is anyway
suppressed due to the vanishing rapidity gap survival 
probability (see Sec.~\ref{sec:cross_section} below).
\subsection{Combining hard and soft interactions}
\label{subsec:combining}
In the second step, we formalize the interplay of hard and soft 
interactions in the amplitude of the hadronic diffractive process, 
Eq.~(\ref{exclusive_diffraction}). To this end, we invoke
the parton picture of the proton wavefunction, as developed 
by Gribov \cite{Gribov:1973jg}. We consider the process 
(\ref{exclusive_diffraction}) in the CM frame, in which the
two protons in the initial state have longitudinal momenta 
$\pm \sqrt{s}/2$, and zero transverse momentum, 
$\bm{p}_{1\perp}, \bm{p}_{2\perp} = 0$. Since $m_H \ll \sqrt{s}/2$, 
angular momentum conservation implies that the reaction amplitude 
is approximately diagonal in the 
transverse coordinates of the colliding protons (\textit{i.e.}, in impact 
parameter) as in two--body elastic scattering at high energies. We thus 
consider partonic configurations centered around the transverse centers of 
the two protons, in which the partons carry fractions of the 
longitudinal momentum of the respective proton.
We may regard the hard scattering process as 
an operator in the basis of these partonic states, denoted by 
$\hat V_{\text{hard}}$. Soft interactions, which build up the partonic 
wavefunctions, are governed by a soft Hamiltonian, $\hat H_{\text{soft}}$. 
While we do not know their explicit form, we can nevertheless state some 
important properties of these operators:
\begin{enumerate}
\item $\hat V_{\text{hard}}$ is local in time (instantaneous) 
on the typical timescale of soft interactions,
\item $\hat V_{\text{hard}}$ is local in transverse position
on the distance scale over which the transverse spatial distribution
of partons in the protons changes due to soft interactions,
\item $\hat V_{\text{hard}}$ preserves the number of partons,
since scattering of both gluons from the same parton dominates in the
hard regime.
\item $\hat V_{\text{hard}}$ preserves the helicity of the colliding partons,
because in a perturbative gauge theory the dominant contribution to the
interaction of partons over large rapidity intervals comes from 
the parton helicity--conserving component of the propagator of 
the exchanged gluon \cite{Gribov:2003nw}.
\end{enumerate}
As a consequence of these properties, we conclude that the operator of 
the hard reaction commutes with the Hamiltonian of soft interactions,
\beq
[ \hat V_{\text{hard}}, \, \hat H_{\text{soft}} ] \;\; = \;\; 0.
\label{hard_soft_commute}
\eeq
This is the mathematical statement of the independence of hard and soft
interactions in diffraction. A schematic illustration of this picture is
given in Fig.~\ref{fig:hardsoft}. We shall refer to 
Eq.~(\ref{hard_soft_commute}) and the formulas derived from it
as the independent interaction approximation.

The amplitude for the double--gap exclusive diffractive process 
(\ref{exclusive_diffraction}) is then determined by the 
matrix element
\beq
T_{\text{diff}} \;\; = \;\;
\langle p_1' p_2'| 
\; \hat S_{\text{soft}} (\infty, 0) \; \hat V_{\text{hard}} \; 
\hat S_{\text{soft}} (0, -\infty) \;
| p_1 p_2 \rangle ,
\eeq
where 
\beq
\hat S_{\text{soft}} (t_2, t_1) \;\; \equiv \;\; 
\text{T} \; \int_{t_1}^{t_2} dt \, \exp (i t \hat H_{\text{soft}} )
\eeq
is the time evolution operator due to soft interactions (we have put the
time of the hard interaction at $t = 0$). Because of 
Eq.~(\ref{hard_soft_commute}), the operator $\hat V_{\text{hard}}$ commutes
with the soft time evolution operator. Using the property
\beq
\hat S_{\text{soft}} (\infty, 0) \, \hat S_{\text{soft}} (0, -\infty) 
\;\; = \;\;
\hat S_{\text{soft}} (\infty, -\infty) 
\;\; \equiv \;\; \hat S_{\text{soft}} ,
\eeq
where $\hat S_{\text{soft}}$ is the $S$--matrix due to soft 
interactions, we obtain
\beq
T_{\text{diff}}
\;\; = \;\; 
\langle p_1' p_2' | \; \hat V_{\text{hard}} \; \hat S_{\text{soft}} \; 
| p_1 p_2 \rangle .
\label{RGS_master}
\eeq
Thus, the amplitude is expressed in terms of the observable matrix 
elements of the soft $S$--matrix, and the operator $\hat V_{\text{hard}}$, 
calculable in QCD at the partonic level. Eq.~(\ref{RGS_master})
is the fundamental expression for discussing the physics 
of RGS within our approach.

\subsection{Suppression of inelastic diffraction}
\label{subsec:inelastic}
In the next step, we evaluate the amplitude for the 
double--gap exclusive diffractive process based on Eq.~(\ref{RGS_master}),
by inserting ``intermediate'' states (actually, states at $t = \infty$)
between the operators. In principle, one needs to sum over all
diffractive states (elastic and inelastic) produced by the 
operator $\hat V_{\text{hard}}$. An important question is which states can
give large contributions to the matrix element. In fact, it turns out 
that the different preferences of hard and soft interactions severely 
restrict the range of states which can effectively contribute. 

Simple arguments show that large--mass diffractive states should
make a negligible contribution to Eq.~(\ref{RGS_master}). If the
two hard gluons in the hard interaction are attached to two different
partons in the proton, the inelastic states predominantly produced
are two jets and gluon bremsstrahlung. It is virtually impossible
to produce such states in soft interactions, hence they cannot
contribute to Eq.~(\ref{RGS_master}). If the two hard gluons are attached 
to the same parton, the cross section of inelastic diffraction is small 
for small $t$ because of the small overlap integral with the inelastic 
state (most of the overlap is with the elastic state), while for large $t$ 
one produces a single parton with transverse momentum 
$p_\perp \sim \sqrt{-t}$, which again
is a state difficult to reach through soft interactions. In addition,
for $t \neq 0$ soft diffraction at LHC energies is known to be dominated 
by the spin--flip amplitude, which further suppresses the overlap integral. 
Together, this restricts the possible mass range of diffractively produced 
states to $M^2_{\text{diff}} \sim \text{few GeV}^2$. 

For a more quantitative estimate, we suppose that the state produced
through inelastic diffraction has the form 
$|pp\rangle + \epsilon | pX \rangle$, where the state $X$ 
is different from the proton, and $\epsilon$ is a small correction.
We can then estimate $\epsilon$ from the Schwarz inequality:
\beq
\frac{\epsilon}{2} \;\; = \;\; 
\sqrt \frac{ \sigma_{\text{soft}}(pp\to Xp) \; 
\sigma_{\text{hard}}(pp\to Xp)}
{\sigma_{\text{soft}}(pp\to pp) \; 
\sigma_{\text{hard}}(pp\to pp)} ,
\eeq
where $\sigma_{\text{hard}}(pp\to pX)$ is the cross section for the 
production of the state $| pX \rangle$ by the operator 
$\hat V_{\text{hard}}$.
Analysis of the Tevatron data (for a review, see 
Ref.~\cite{Goulianos:2005ac})
shows that the fraction of diffractive events in soft collisions decreases 
with increasing energy, and that the distribution over the excitation mass
is $\propto 1/M_X^2$. As a result, we expect that at the LHC energy 
$\epsilon \le 2 \cdot  10^{-2}$. Thus, the diffractively produced state 
is actually the $|pp\rangle$ state, and the contributions from
inelastic diffraction are small. 

The small overlap between hard and soft diffraction can also be understood
as the result of the different impact parameter dependence of both types 
of processes. Hard diffraction occurs mostly at small
impact parameters, $b^2 \sim B_g$. Soft diffraction, because of the
approach to the BDL, occurs mostly at large impact parameters, 
$b^2 \sim B$, which, moreover, rapidly grow with the collision energy. 
We note again that the peripheral nature of soft diffraction
was established already within Reggeon field theory, where it was
found that the BDL solves the consistency problem associated with the
triple Reggeon formula \cite{Marchesini:1976hw}.

The restriction to the $pp$ intermediate state turns Eq.~(\ref{RGS_master})
into a tool for quantitative evaluation of the diffractive amplitude
and the RGS. In particular, with the $pp$ intermediate state we
can approximate the matrix element of the soft--interaction 
time evolution operator by that of the full $S$--matrix, 
\textit{i.e.}, the $pp$ elastic scattering amplitude, which
is known experimentally; see Section~\ref{sec:transverse_soft}.
For this approximation to be legitimate it is crucial that
scattering at small impact parameters turns out to be strongly suppressed
due to the approach to the BDL in $pp$ elastic scattering,
as will be seen from the results of Section~\ref{sec:cross_section}.
The diffractive process is thus dominated by large impact parameters,
where $pp$ elastic scattering is dominated by soft interactions.

In the studies of double--gap exclusive diffraction based on the
pomeron model of Ref.~\cite{Khoze:2000wk}, inelastic intermediate 
states were effectively included by way of a two--component formalism
(however, no explicit non-diagonal ``transition'' GPDs were introduced).
We have argued here that these contributions are strongly suppressed,
because of the small overlap of states accessible in hard and soft
interactions. We shall comment on the implications of this for
the numerical values of the RGS probability 
in Section~\ref{subsec:numerical}

%
%
\begin{figure*}
\includegraphics[width=14cm]{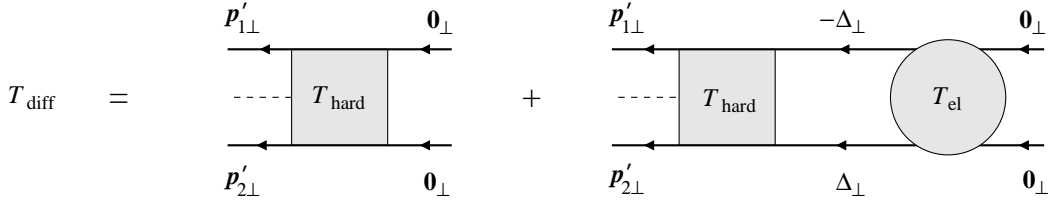} 
\caption{The amplitude for double--gap exclusive hard diffraction in momentum 
representation, Eqs.~(\ref{amp_momentum})--(\ref{amp_momentum_t}).
The first term is the amplitude of the hard reaction alone,
the second term the correction due to soft elastic rescattering.
Only the transverse momenta of the protons are indicated;
the momentum transfer due to soft elastic scattering is 
$\bm{\Delta}_\perp$.}
\label{fig:mom}
\end{figure*}
\subsection{Evaluation of the diffractive amplitude}
\label{subsec:evaluation}
It remains to actually evaluate the matrix element (\ref{RGS_master})
with $|pp\rangle$ intermediate states, using the specific form
of the hard scattering amplitude and the $pp$ elastic
scattering amplitude. We insert a set of $pp$ intermediate
states in the form 
\beq
\int\! \frac{d^3 p_1^{''}}{(2\pi )^3 \sqrt{s}} \;
\int\! \frac{d^3 p_2^{''}}{(2\pi )^3 \sqrt{s}} \; 
|p_1^{''} p_2^{''} \rangle \langle p_1^{''} p_2^{''} | ,
\eeq
where we have approximated the energy of the individual protons 
by $\sqrt{s}/2$. The matrix element of the operator 
$\hat V_{\text{hard}}$ between the two--proton states is, 
by definition, given by [\textit{cf.}\ Eqs.~(\ref{A_hard_GPD})]
\be
\langle p_1^{'} p_2^{'} | \; \hat{V}_{\text{hard}} \; |  p_1^{''} p_2^{''} 
\rangle
&=& \kappa (s, \xi_1, \xi_2 ) \nonumber \\
&\times& F_g (x_1, \xi_1, \tilde t_1; Q^2) \nonumber \\
&\times& F_g (x_2, \xi_2, \tilde t_2; Q^2) ,
\ee
where 
\be
\tilde t_1 &\equiv&
-(\bm{p}_{1\perp}^{'} - \bm{p}_{1\perp}^{''})^2 ,
\\
\tilde t_2 &\equiv&
-(\bm{p}_{2 \perp}^{'} - \bm{p}_{2 \perp}^{''})^2 .
\ee
The factor
\be
\kappa (s, \xi_1, \xi_2 ) &\equiv& C_{\text{hard}} \nonumber \\
&\times& H_g (x_1, \xi_1, t_1 = 0) \nonumber \\
&\times& H_g (x_2, \xi_2, t_2 = 0) 
\label{kappa}
\ee
represents the symbolic result for the absolute normalization of amplitude 
of the hard scattering process; it contains the amplitude of the two--gluon 
exchange process, $C_{\text{hard}}$, including the information about the 
$ggH$ coupling given by the electroweak theory, as well as the information 
about the gluon GPD in the colliding protons at $t = 0$. 
The information about the transverse momentum dependence of the amplitude 
is contained in the two--gluon formfactors, $F_g$,
\textit{cf.}\ Eq.(\ref{twogl_def}). Furthermore, we replace
in Eq.~(\ref{RGS_master})
\beq
\hat S_{\text{soft}} 
\;\; \rightarrow \;\; \hat S \;\; = \;\; 1 \; + \; (\hat S - 1), 
\label{S_decomposition}
\eeq
and use the fact that the matrix element of the operator 
$\hat S - 1$ is given by
\be
\langle p_1^{''} p_2^{''} | \; \hat{S} - 1 \; |  p_1 p_2 \rangle
&=& i (2\pi)^4 \; \delta^{(4)} (p_1^{''} + p_2^{''} - p_1 - p_2)
\nonumber \\
&\times& (4\pi) \, T_{\text{el}} (s, t) ,
\label{S_minus_1_me}
\ee
with
\beq
t \;\; = \;\; - (\bm{p}_{1\perp}^{''} - \bm{p}_{1\perp})^2 
  \;\; = \;\; - (\bm{p}_{2\perp}^{''} - \bm{p}_{2\perp})^2 .
\label{S_minus_1_t}
\eeq
Finally, taking into account that at high energies the 
energy--conserving delta function in Eq.~(\ref{S_minus_1_me})
effectively conserves longitudinal momentum, and combining the
contributions from the two terms in Eq.~(\ref{S_decomposition}),
we obtain 
\be
\lefteqn{T_{\text{diff}} (\bm{p}_{1\perp}', \bm{p}_{2\perp}')
\;\; = \;\; \int\frac{d^2 \Delta_\perp}{(2\pi)^2}} &&
\nonumber \\
&\times& \kappa \; F_g \left( x_1, \xi_1, \tilde t_1; \, Q^2 \right) \;
F_g \left( x_2, \xi_2, \tilde t_2; \, Q^2 \right) 
\nonumber \\
&\times & \left[ (2\pi)^2 \delta^{(2)}(\bm{\Delta}_\perp)
\; + \; \frac{4\pi i}{s} T_{\text{el}} (s, t) 
\right] ,
\label{amp_momentum}
\ee
where now
\be
\tilde t_1 &\equiv& -(\bm{p}_{1\perp}' - \bm{\Delta}_\perp)^2 , \, 
\\
\tilde t_2 &\equiv& -(\bm{p}_{2\perp}' + \bm{\Delta}_\perp)^2 , 
\\
t &\equiv& -\bm{\Delta}_\perp^2 .
\label{amp_momentum_t}
\ee
This result has a simple interpretation (see Fig.~\ref{fig:mom}). 
The first term in the
brackets represents the amplitude of the hard reaction alone.
The second term represents the contribution in which the 
hard reaction is accompanied by soft elastic rescattering
with transverse momentum transfer $\bm{\Delta}_\perp$.
The total amplitude is the coherent superposition of 
the two contributions. We note that the form of this result
is analogous to the well--known absorption corrections 
in Regge phenomenology, in which an elementary Regge pole amplitude
is modified by elastic rescattering.

\begin{figure}
\includegraphics[width=5cm]{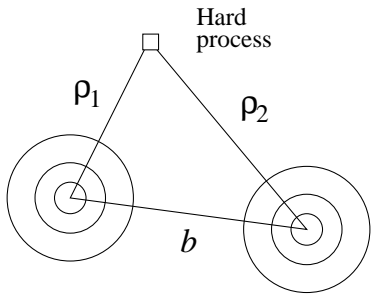} 
\caption{Illustration of the transverse coordinate representation
of the diffractive amplitude, 
Eq.~(\ref{T_rho}). The hard scattering process is local in transverse space.
The centers of the colliding protons are displaced by the distance
$b = |\bm{b}|$, and $\bm{\rho}_{1, 2} = |\bm{\rho}_{1, 2}|$ are the 
distances from the centers to the point of the hard process.}
\label{fig:coord}
\end{figure}

It is instructive to convert the result (\ref{amp_momentum}) to 
the transverse coordinate representation. 
Substituting the Fourier representation
of the gluon GPDs, Eq.~(\ref{rhoprof_def}), and the representation
of the $pp$ elastic amplitude in terms of the profile function,
Eq.~(\ref{Gamma_def}), and using standard Fourier transform
manipulations, we obtain
\be
\lefteqn{T_{\text{diff}} (\bm{p}_{1\perp}', \bm{p}_{2\perp}')
\;\; = \;\; \int d^2 b \int d\rho_1 \int d\rho_2}
&& \nonumber \\
&\times&  \delta^{(2)} (\bm{b} - \bm{\rho}_1 + \bm{\rho}_2)
\; e^{-i (\bm{p}_{1\perp}' \bm{\rho}_1) 
-i (\bm{p}_{2\perp}' \bm{\rho}_2)} 
\nonumber \\
&\times& \kappa \; 
F_g \left( x_1, \xi_1, \bm{\rho}_1; \, Q^2 \right) 
\; F_g \left( x_2, \xi_2, \bm{\rho}_2; \, Q^2 \right) 
\nonumber \\
&\times& 
\left[ 1 - \Gamma(s, \bm{b}) \right] .
\label{T_rho}
\ee
Here the scattering amplitude is represented as the coherent superposition 
of amplitudes corresponding to $pp$ scattering at given transverse
displacement (impact parameter), $\bm{b}$. 
The amplitude for the hard process is proportional
to the product of the transverse spatial gluon transition densities
at positions $\bm{\rho}_{1, 2}$ relative to the centers of the 
respective protons, with the three transverse vectors satisfying
the triangular condition $\bm{\rho}_1 - \bm{\rho}_2 = \bm{b}$
(see Fig.~\ref{fig:coord}). The modifications due to elastic 
rescattering now take the 
form of a multiplication of the hard scattering amplitude with
the ``absorption factor,'' $1 - \Gamma(s, b)$. Note that
the modulus squared of this factor can be interpreted as the
probability for ``no inelastic interaction'' in $pp$ scattering
at a given impact parameter, \textit{cf.}\ Eq.~(\ref{P_noin}).
This interpretation will be explored further in Sec.~\ref{sec:cross_section}.
%
%

Our partonic approach allows us to calculate the amplitude for double--gap
exclusive diffraction in a model--independent way in terms of the gluon 
GPD and the phenomenological $pp$ elastic scattering amplitude; see 
Eqs.~(\ref{amp_momentum}--\ref{T_rho}). In Ref.~\cite{Kaidalov:2003fw} 
such processes were studied using a model of elastic $pp$ scattering 
which included the enhanced eikonal series of single Pomeron exchanges 
and the triple--Pomeron vertex to describe the soft spectator interactions.
The expression for the amplitude in the case of a single Pomeron exchange
in that model (and without inelastic intermediate states) would formally 
coincide with our expressions (\ref{amp_momentum}--\ref{T_rho}). 
Whether the same is true for the full amplitude in that model is 
less clear; \textit{cf.}\ the discussion of the numerical results 
in Section~\ref{subsec:numerical} below. 
\section{The rapidity gap survival probability}
\label{sec:cross_section}
We now use our general result for the amplitude of double--gap exclusive 
diffractive processes in the independent interaction approximation, 
Eq.~(\ref{T_rho}), to calculate the RGS probability for such processes. 
At this level of approximation, we shall recover a simple ``geometric'' 
expression for the RGS probability, which permits a probabilistic 
interpretation and was heuristically derived in 
Refs.~\cite{Frankfurt:2004kn,Frankfurt:2004ti,Frankfurt:2005mc}.
We discuss the impact parameter dependence of RGS and stress the
the crucial role of the BDL in stabilizing the numerical estimates
ensuring a model--independent result.
\subsection{Impact parameter representation}
\label{subsec:geometry}
In order to compute the cross section for double--gap exclusive diffractive 
production of a given state at fixed rapidity, we integrate the 
modulus squared of the amplitude (\ref{T_rho}) over the
final proton transverse momenta. By standard Fourier transform 
manipulations we obtain
\be
\sigma_{\text{diff}} &=& \text{(kinematic factors)} \; \times 
\int \frac{d^2 p_{1\perp}'}{(2\pi )^2} \; 
\int \frac{d^2 p_{2\perp}'}{(2\pi )^2} \; 
\nonumber \\
&\times & 
\left| T_{\text{diff}} (\bm{p}_{1\perp}', \bm{p}_{2\perp}') \right|^2
\\
&=& \text{(const)} \; \times \; \int d^2 \rho_1 \; \int d^2 \rho_2 \; 
F^2_g (\bm{\rho}_1) \; F^2_g (\bm{\rho}_2) \; 
\nonumber \\
&\times & 
\left| 1 - \Gamma (\bm{\rho}_2
- \bm{\rho}_1) \right|^2 
\label{sigma_full}
\ee
(for brevity we suppress all arguments except the transverse 
coordinates in $F_g$ and $\Gamma$).
The RGS probability due to soft interactions \cite{Bjorken:1992er}, 
by definition, is given by the ratio of the cross section of the 
physical double--gap 
diffractive process (\ref{sigma_full}) to the cross section of the 
hypothetical process with the same hard scattering subprocess but 
with no soft spectator interactions, corresponding to 
expression (\ref{sigma_full}) with $\Gamma \equiv 0$, 
\beq
S^2 \;\; \equiv \;\; 
\frac{\sigma_{\text{diff}} \, (\text{physical}) \phantom{xxi}}
{\sigma_{\text{diff}} \, (\text{no soft interactions})} .
\label{survb_def}
\eeq
We can cast this ratio in a simple form. We rewrite 
the convolution integral in Eq.~(\ref{sigma_full}) by inserting 
unity in the form (\textit{cf.}\ Fig.~\ref{fig:coord})
\beq
\int d^2 b \; \delta^{(2)} (\bm{b} + \bm{\rho}_1 - \bm{\rho}_2 ) ,
\eeq
and introduce a normalized impact parameter distribution,
\be
\lefteqn{
P_{\text{hard}} (\bm{b}) \;\; \equiv \;\;
\int d^2 \rho_1 \; \int d^2 \rho_2 \; 
\delta^{(2)} (\bm{b} + \bm{\rho}_1 - \bm{\rho}_2 ) }
&& \nonumber \\
&\times& 
\frac{F^2_g (\bm{\rho}_1)}
{\displaystyle \left[ \int d^2 \rho_1' \; F^2_g (\bm{\rho}_1') \right]}
\; \frac{F^2_g (\bm{\rho}_2)}
{\displaystyle \left[ \int d^2 \rho_2' \; F^2_g (\bm{\rho}_2') \right]} ,
\label{P_hard}
\ee
which satisfies
\beq
\int d^2 b \; P_{\text{hard}} (\bm{b}) \;\; = \;\; 1 .
\label{P_hard_normalization}
\eeq
In terms of this distribution the RGS probability (\ref{survb_def})
is expressed as
\beq
S^2 \;\; = \;\; 
\int d^2 b \; P_{\text{hard}} (\bm{b}) \; |1 - \Gamma (\bm{b})|^2 .
\label{survb}
\eeq
This result agrees with the expression for the RGS probability derived 
heuristically in 
Refs.~\cite{Frankfurt:2004kn,Frankfurt:2004ti,Frankfurt:2005mc}
\footnote{Note that the distribution $P_{\text{hard}} (\bm{b})$ of 
Eq.~(\ref{P_hard}) (overlap integral of squared spatial distribution
of gluons) is different from the distribution $P_4(\bm{b})$
(square of overlap integral of spatial distribution of gluons), 
which was introduced in Ref.~\cite{Frankfurt:2003td} to describe 
double dijet production
in inclusive high--energy $pp / \bar p p$ scattering. The two distributions
coincide only in the case of an exponential parametrization of the two--gluon
formfactor, which corresponds to a Gaussian dependence of the 
formfactors on transverse momenta and coordinates; see 
Eqs.~(\ref{spatial_gaussian}) and (\ref{P_hard_gaussian}). 
In general, it is not correct to replace $P_{\text{hard}} (\bm{b})$ 
by $P_4(\bm{b})$, as was done in 
Refs.~\cite{Frankfurt:2004kn,Frankfurt:2004ti}.}.
For the comparison of our result for the RGS probability 
with that obtained with the pomeron model of Ref.~\cite{Kaidalov:2003fw} 
we refer to Section~\ref{subsec:numerical} below; 
see also the comments at the end of Sec.~\ref{subsec:evaluation}.

Expression~(\ref{survb}) for the RGS probability 
allows for a simple probabilistic interpretation. Consider a $pp$ collision 
at given impact parameter, $b = |\bm{b}|$. Since the hard two--gluon
exchange process is effectively local in transverse space, the probability 
for it to happen is proportional to the product of the 
squared transverse spatial distributions of gluons in the two colliding 
protons, integrated over the transverse plane,
as given by the numerator of Eq.~(\ref{P_hard}). 
Consider now a hypothetical sample of $pp$ events with the 
two--gluon induced hard scattering process, 
but an otherwise arbitrary (non-diffractive) final state. By the laws of 
probability, the distribution of impact parameters in this sample is given
by the normalized distribution $P_{\text{hard}}(\bm{b})$, Eq.~(\ref{P_hard}). 
A diffractive event results if the spectator systems of the two protons 
do not interact inelastically. The probability for this to happen 
in a $pp$ collision at fixed $b$ is given by $|1 - \Gamma (\bm{b})|^2$,
\textit{cf.}\ Eq.~(\ref{P_noin}),
in analogy to the well--known formula for inelastic scattering in
non-relativistic theory \cite{LLIII}. The RGS probability, 
which is defined as the fraction of diffractive events in the sample of 
all events containing the same hard scattering process, is then
given by the average of this function with the normalized 
$b$--distribution in the sample, Eq.~(\ref{survb}).

It needs to be stressed that the impact parameter of a single 
$pp$ event is not observable, being a microscopic quantity beyond the
reach of any experimental apparatus. In the above arguments, the
impact parameter plays the role of a randomly chosen external parameter.
However, using information about the transverse spatial distribution 
of gluons in the proton from independent measurements, we can calculate 
the probability for certain hard processes in a $pp$ collision as a 
function of the impact parameter, and thus infer the distribution
of impact parameters in a sample of events with the
same hard process. This logic was used in Ref.~\cite{Frankfurt:2003td}
to devise a trigger on central collisions in inclusive $pp$ scattering
by requiring hard dijet production at small rapidities. Here we
use the same strategy to model soft spectator interactions
in double--gap exclusive diffractive $pp$ scattering.

%
%
\begin{figure}
\includegraphics[width=8cm]{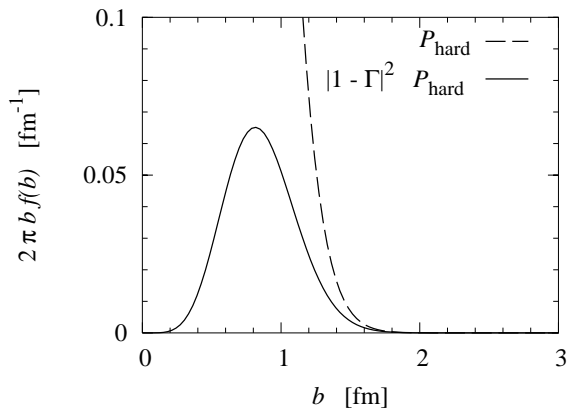}
\caption[]{The integrand (impact parameter distribution) 
in the RGS probability, Eq.~(\ref{survb}), 
for Higgs boson production at the LHC energy. Dashed line: 
$b$--distribution of the hard two--gluon exchange, $P_{\text{hard}} (b)$, 
Eq.~(\ref{P_hard}), evaluated with the 
exponential parametrization of the two--gluon formfactor,
Eq.~(\ref{twogl_exponential}) with $B_g = 3.24\, \text{GeV}^{-2}$.
Solid line: The product $P_{\text{hard}} (b) |1 - \Gamma (\bm{b})|^2$,
evaluated with the exponential parametrization, 
Eq.~(\ref{Gamma_gaussian}), with $B = 21.8 \, \text{GeV}^{-2}$. 
The vanishing of $|1 - \Gamma (\bm{b})|^2$,
at small $b$, \textit{cf.}\ Fig.~\ref{fig:gamma}, strongly suppresses
contributions from small impact parameters. Note that the plot 
shows $2\pi b$ times the functions of impact parameter; the 
small--$b$ part of the dashed curve [distribution $P_{\text{hard}} (b)$] 
would be close to the left boundary of the plot and was omitted for 
better legibility. The RGS probability, $S^2$, 
is given by the area under the solid curve.}
\label{fig:survb}
\end{figure}
The integrand in Eq.~(\ref{survb}) describes the effective distribution
of impact parameters in a sample of double--gap diffractive events, 
and reflects the interplay of hard and soft interactions at the
cross section level. The probability for the hard process,
$P_{\text{hard}}(\bm{b})$, favors small impact parameters, which maximize
the overlap of the large--$x$ gluon distributions in the protons, 
and vanishes for $b^2 \gg 1/B_g$. The probability for no inelastic soft
interactions, $|1 - \Gamma (\bm{b})|^2$, favors large impact
parameters, which increase the chances for the protons to stay intact,
and vanishes for $b^2 \ll 1/B$ where $pp$ scattering approaches the 
BDL. The product of the two probabilities 
is suppressed both at small and at large $b$, 
and thus concentrated at intermediate values of $b$. 

This point can nicely be illustrated with the Gaussian
parametrizations of the transverse spatial distribution of gluons, 
Eq.~(\ref{spatial_gaussian}), and the $pp$ elastic profile function, 
Eq.~(\ref{Gamma_gaussian}). With the Gaussian form (\ref{spatial_gaussian}),
the convolution integral in Eq.~(\ref{P_hard}) can be computed 
analytically,
\beq
P_{\text{hard}} (\bm{b}) \;\; = \;\; 
\frac{\exp\left[ -\bm{b}^2 / (2B_g) \right]}{2 \pi B_g} .
\label{P_hard_gaussian}
\eeq
This function is shown by the dashed line in Fig.~\ref{fig:survb}. 
The integrand of Eq.~(\ref{survb}) is given by
\be
\lefteqn{P_{\text{hard}} (\bm{b}) \; |1 - \Gamma (\bm{b})|^2} &&
\nonumber \\
&=& \frac{1}{2 \pi B_g} \exp \left( -\frac{\bm{b}^2}{2B_g} \right)
\; \left[ 1 - \exp \left( -\frac{\bm{b}^2}{2B} \right) \right]^2 ,
\label{combined_gaussian}
\ee
and is shown by the solid line in Fig.~\ref{fig:survb}.
It is suppressed both for $b^2 \ll 1/B$ (because of the ``blackness'' of 
the $pp$ amplitude) and for $b^2 \gg 1/B_g$ (because of the vanishing of 
the overlap of the two gluon distributions), and thus concentrated at 
intermediate values of $b$. The maximum of $2\pi b$ times
the combined distribution is at
\beq
b^2 \;\; \approx \;\; 5 B_g \hspace{3em} (B_g \ll B) .
\label{b2_max}
\eeq
We see that within our two--scale picture of the transverse structure
of hard and soft interactions, \textit{cf.}\ Fig.~\ref{fig:twoscale}, the 
dominant impact parameters in double--gap exclusive diffractive processes
are determined by $B_g$ --- the smaller of the 
two areas ---, but may be numerically large because of a 
large numerical factor. The RGS probability,
Eq.~(\ref{survb}), is given by the integral of $2\pi b$ times
Eq.~(\ref{combined_gaussian}) (\textit{i.e.}, the area under the
solid curve in Fig.~\ref{fig:survb}), and can be computed
analytically,
\beq
S^2 \;\; = \;\; \frac{2 B_g^2}{(B + B_g) (B + 2 B_g)} 
\;\; \approx \;\; \frac{2 B_g^2}{B^2} \hspace{3em} (B_g \ll B) .
\label{S2_exponential_gen}
\eeq
The gap survival probability is of the order $(B_g / B)^2$,
\textit{i.e.}, it is proportional to the square of the ratio
of the transverse area occupied by hard gluons to the area 
corresponding to soft interactions. Thus, our two--scale picture
offers a parametric argument for the smallness of the rapidity gap 
survival probability. 

The approach to the black--disk limit in $pp$ scattering at 
high energies, \textit{i.e.}, the fact that 
$\Gamma (\bm{b}) \rightarrow 1$ at small $b$, plays a crucial
role in determining the numerical value of the RGS probability
and ensuring stability of our calculation with respect to 
variation of the parameters.
A small deviation of the profile function from unity at $b = 0$,
of the form $\Gamma (\bm{b} = 0) = 1 - \epsilon$ with $\epsilon \ll 1$, 
would change the result for the gap survival probability to
\beq
S^2 \;\; \rightarrow \;\; S^2|_{\text{BDL}} \; + \; \epsilon^2 
\eeq
[here we have used that $B_g \ll B$, and that the integral of 
$P_{\text{hard}}$ is unity, \textit{cf.}\ Eq.~(\ref{P_hard_normalization})].
The approach to the BDL effectively eliminates $\Gamma (\bm{b} = 0)$ as a
parameter in our calculation. We stress again that the experimental 
data as well as theoretical arguments indicate that the BDL is indeed
reached in $pp$ scattering at small impact parameters at the LHC energy. 
\subsection{Numerical estimates}
\label{subsec:numerical}
%
%
\begin{figure}
\includegraphics[width=8cm]{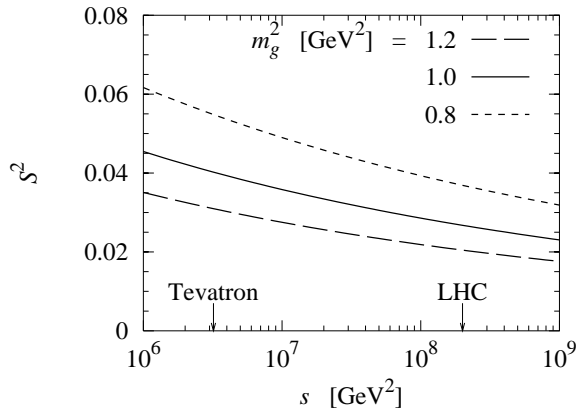}
\caption[]{The RGS probability for double--gap exclusive diffractive processes,
Eq.~(\ref{survb}), as a function of the squared CM energy, $s$. 
The Tevatron and LHC energies are marked by arrows. 
Shown are the results obtained with the dipole parametrization
of the two--gluon formfactor (\ref{twogl_dipole}), for different 
values of the mass parameter, $m_g^2$. The value $m_g^2 = 1 \, \text{GeV}^2$
is appropriate for Higgs boson production at the LHC at central rapidities.
The profile function of the $pp$ elastic amplitude was taken from 
Ref.~\cite{Islam:2002au} (\textit{cf.}\ Fig.~\ref{fig:gamma}).}
\label{fig:s2_sdep}
\end{figure}
For a numerical estimate of the gap survival probability we 
evaluate Eq.~(\ref{survb}) with the 
dipole parametrization of the two--gluon formfactor, 
Eq.~(\ref{twogl_dipole}), and the parametrization of the $pp$
elastic amplitude of Ref.~\cite{Islam:2002au}. For Higgs production
at the LHC ($\sqrt{s} = 14 \, \text{TeV}$)
at central rapidities the momentum fractions of the annihilating 
gluons are $x_{1, 2} \sim 10^{-2}$ (at a scale $Q^2 \ll m_H^2$). 
For such values of $x$ in principle the contributions of the nucleon's
pion cloud to the gluon density at transverse distances 
$\rho \sim 1/(2 M_\pi)$ need to be taken into account; see
Section~\ref{sec:transverse_hard}. As we shall explain 
below, these contributions to the
gluon density involve correlations in the nucleon wavefunction, 
which effectively reduce their contribution to RGS, and should not
be included in the estimate based on Eq.~(\ref{survb}). 
We therefore use in our estimate at the LHC energy the simple dipole 
formfactor with $m_g^2 \approx 1\, \text{GeV}^2$, which does not 
include the pion cloud contribution. With this choice of parameters
Eq.~(\ref{survb}) gives for the RGS probability for Higgs production 
at the LHC
\beq
S^2 \;\; = \;\; 0.027 .
\label{S2_LHC_num}
\eeq
The energy dependence of the RGS probability is shown in 
Fig.~\ref{fig:s2_sdep}, for various values of the mass parameter
of the two--gluon formfactor, $m_g^2$. At the Tevatron energy 
($\sqrt{s} = 1.9 \, \text{TeV}$), the gluon momentum fractions
$x_{1, 2}$ are of the order $\sim 10^{-1}$, for which the pion cloud
contributions to the gluon density are naturally absent. While a mass 
parameter $m_g^2 = 1 \, \text{GeV}^2$ is still reasonable in this situation,
even higher values $m_g^2$ might be relevant in this case. Taking into
account this effective change of $m_g^2$ with $s$ via the 
$x_1, x_2$--dependence, the actual variation of the RGS probability 
between the LHC and the Tevatron energies is less pronounced than 
it appears from the fixed--$m_g^2$ curves in Fig.~\ref{fig:s2_sdep}.

%
%
\begin{figure}
\includegraphics[width=8cm]{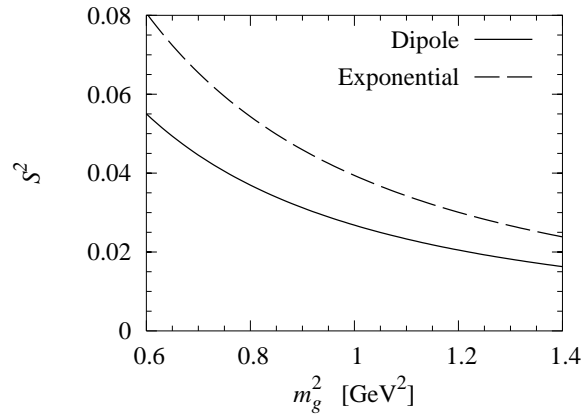}
\caption[]{Dependence of the RGS probability for double--gap exclusive 
diffractive Higgs production at the LHC, Eq.~(\ref{survb}), on the mass 
parameter in the dipole parametrization of the two--gluon formfactor, 
$m_g^2$, Eq.~(\ref{twogl_dipole}). Also shown are the results obtained
with the exponential parameterization, Eq.~(\ref{twogl_exponential}),
with $B_g = 3.24 / m_g^2$, \textit{cf.}\ Eq.~(\ref{dip_exp}) and
Fig.~\ref{fig_fg_dip_exp}. The profile function of the $pp$ elastic 
amplitude in both cases is the one of Ref.~\cite{Islam:2002au}.}
\label{fig:s2_mdep}
\end{figure}
Figure~\ref{fig:s2_mdep} (solid line) shows the dependence of the 
RGS probability at the LHC energy
on the mass parameter in the dipole parametrization of the 
two--gluon formfactor, $m_g^2$, Eq.~(\ref{twogl_dipole}). 
The value of $S^2$ strongly increases with decreasing $m_g^2$, 
\textit{i.e.}, with increasing radius of the 
the transverse spatial distribution of gluons,
similar to the behavior found with the simple exponential 
parametrizations, Eq.~(\ref{S2_exponential_gen}).
To illustrate the sensitivity of our numerical
predictions to the shape of the two--gluon formfactor,
we also show the results for $S^2$ obtained when replacing the
dipole formfactor by the exponential, Eq.~(\ref{twogl_exponential}), 
with $B_g = 3.24 / m_g^2$, \textit{cf.}\ Eq.~(\ref{dip_exp}).
(Figure~\ref{fig:s2_mdep}, dashed line).
One sees that the numerical values are rather different,
in spite of the apparent similarity of the two formfactors over a
wide range, $|t| \lesssim 1 \, \text{GeV}^2$ (\textit{cf.}\ 
Fig.~\ref{fig_fg_dip_exp}). Comparing the two curves of 
Figure~\ref{fig:s2_mdep}, we estimate the uncertainty of our
numerical prediction for the RGS probability due to the uncertainty 
of the shape of the two--gluon formfactor to be at least 
$\sim 30\%$.

In a similar way, we can estimate the uncertainty of the RGS
probability due to the uncertainty of the profile of the $pp$ elastic 
amplitude. Comparing the numerical values obtained from 
Eq.~(\ref{survb}) with the parametrization of Ref.~\cite{Islam:2002au}
and with the simple exponential parametrization, Eqs.~(\ref{A_exponential})
and (\ref{Gamma_gaussian}), for the same two--gluon formfactor, 
we again find variations of the order of $\sim 30\%$.

The relatively high sensitivity of the numerical estimates
to the shape of the two--gluon formfactor and the profile function
of the $pp$ elastic amplitude can be understood as a result of the 
peculiar interplay of hard and soft interactions in Eq.~(\ref{survb}). 
The different impact parameter dependence of hard and soft interactions 
(\textit{cf.}\ Fig.~\ref{fig:survb} and the discussion above) 
essentially eliminates contributions from the 
regions corresponding to the ``bulk'' of the individual distributions, 
$P_{\text{hard}} (\bm{b})$ and $|1 - \Gamma (\bm{b})|$, allowing 
for significant strength only in an intermediate region of
impact parameters, where there is considerable sensitivity to
the shape of the distributions. This seems to be a principal 
feature of estimates of the RGS probability based on Eq.~(\ref{survb}). 

Detailed numerical studies of the RGS probability were
made within the eikonalized Pomeron model for soft interactions, 
see Ref.~\cite{Khoze:2000wk} for a summary. We would like to comment 
on these estimates from the perspective of our approach. As we 
already noted, the approach of Ref.~\cite{Khoze:2000wk} includes
contributions from inelastic intermediate states (albeit without
introducing non-diagonal ``transition'' GPDs). We have argued that 
within the independent interaction approximation these contributions 
are strongly suppressed and should not be included, 
see Section~\ref{subsec:inelastic}. Nevertheless, the numerical
result for the RGS probability in Higgs boson production in 
double--gap diffraction at the LHC quoted in Ref.~\cite{Kaidalov:2003fw},
$S^2 = 0.023$ \footnote{The value $S^2 = 0.026$ 
quoted in Ref.~\cite{Kaidalov:2003fw} is obtained when taking into
account corrections to the hard scattering amplitude resulting from 
the proton transverse momenta, which are not included in our approach.
The difference between the two values is immaterial for the present
discussion.}, is rather similar to our estimate (\ref{S2_LHC_num}). 
Note that Ref.~\cite{Kaidalov:2003fw} ascribes an uncertainty of
$\sim 50\%$ to this value; we have estimated a similar uncertainty for
our result due to the combined uncertainties in the profile function 
and the two--gluon formfactor (see above). It is interesting to ask 
why the results are so similar when the two approaches differ in
their treatment of inelastic diffraction. In order to clarify this
question, we have evaluated our expression for the RGS 
probability, Eq.~(\ref{survb}), with the profile function of the $pp$ 
elastic amplitude obtained within the model of Ref.~\cite{Khoze:2000wk}, 
which was kindly provided to us by M.~Ryskin; we emphasize that this is 
not the same as evaluating the expression for the RGS probability given in 
Ref.~\cite{Khoze:2000wk}. Using the exponential form of the two--gluon
formfactor (\ref{twogl_exponential}) with parameters
$B_g = (4, \, 5.5, \, 10.1 ) \, \text{GeV}^{-2}$ 
we obtain in this way $S^2 = (0.042, \, 0.069, \, 0.157)$, which should be 
compared to the results quoted in Ref.~\cite{Khoze:2000wk}, 
$S^2 = (0.02, \, 0.04, \, 0.11)$. One sees that our Eq.~(\ref{survb})
gives systematically larger values than the approach of 
Ref.~\cite{Khoze:2000wk} for the same profile function and the same 
two--gluon formfactor. This difference should be attributed to the
effect of inelastic diffraction. What is then quoted as the final estimate 
of $S^2$ depends on the preferred value of $B_g$. The RGS probability strongly 
decreases with $B_g$, \textit{cf.}\ Eq.~(\ref{S2_exponential_gen}) 
and Fig.~\ref{fig:s2_mdep}. For Higgs production at the LHC we use a value 
of $B_g = 3.24 \, \text{GeV}^{-2}$ (corresponding to a mass 
parameter in the dipole parametrization of $m_g^2 = 1\,\text{GeV}^2$), 
which is based on analysis of the $J/\psi$ photoproduction data over a 
wide range of energies and takes into account the effects of QCD evolution
(\textit{cf}.\ Section~\ref{sec:transverse_hard}). 
With this value we obtain $S^2 = 0.030$ for the exponential two-gluon 
formfactor and the profile function of Ref.~\cite{Khoze:2000wk}.
This value of $B_g$ is lower than the range of values considered in
Ref.~\cite{Khoze:2000wk} (our $B_g = 2 b$ in the notation of that paper). 
The value of $B_g$ for Higgs production at the LHC taken in 
Ref.~\cite{Kaidalov:2003fw} is $B_g = 4 \, \text{GeV}^{-2}$, 
which results in $S^2 = 0.02$ in their model. One sees that the different 
values of $B_g$ partly compensate the differences due to the 
treatment of inelastic diffraction in the two approaches. 
We thus conclude that the similarity of the final numerical 
estimate of Ref.~\cite{Kaidalov:2003fw} with our results
is somewhat accidental. In any case, the differences between 
the numerical results of both approaches are within the 
estimated uncertainties.

Potentially more important than the uncertainties of our calculation
of the RGS probability within the independent interaction approximation 
are effects of possible correlations between hard and soft interactions. 
These effects can naturally be incorporated into our partonic picture,
and further decrease the RGS probability compared to the
independent interaction approximation 
(see Section~\ref{subsec:correlations}).
\section{Beyond the independent interaction approximation}
\label{sec:beyond}
Our treatment of RGS so far was based on the idea that
hard and soft interactions are approximately independent because
they proceed over widely different time-- and distant scales.
It is clear that this approximation has certain limitations, 
concerning both the range of its applicability and its accuracy.
In this Section we discuss various physical effects which 
violate the assumption of independence of hard and soft interactions
and give rise to corrections to the estimates of the RGS probability
of Section~\ref{sec:cross_section}. These are (a) the increase of 
hard screening corrections with energy, (b) correlations
between hard and soft interactions. These corrections have not been
considered in previous treatments of RGS in 
Refs.~\cite{Kaidalov:2003fw,Khoze:1997dr,Khoze:2000cy}.
\subsection{Hard screening corrections}
The independent interaction approximation relies on the assumption
of widely different characteristic scales of hard and soft processes.
However, the difference between the two scales tends to decrease, 
and may even disappear, at high energies, because of the fast increase 
of the amplitudes of hard processes with energy. 
One example of this effect is the BDL in $pp$ elastic scattering
at central impact parameters, which can be explained both on the basis
of hard and soft interactions (see Section~\ref{sec:transverse_soft}). 
In diffractive scattering, the increase of hard amplitudes at 
high energies leads to a reduction of the RGS probability relative 
to the estimates presented in Section~\ref{sec:cross_section}. 

\begin{figure}
\includegraphics[width=8cm]{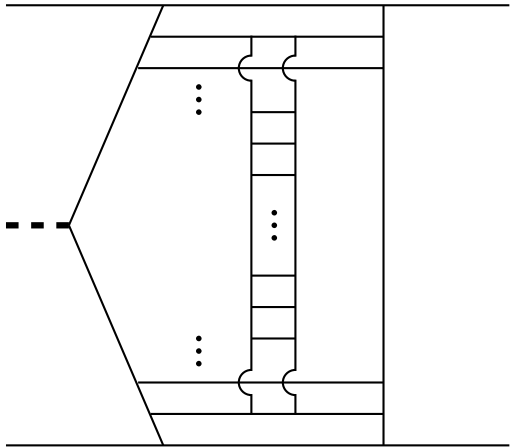}
\caption[]{A typical diagram describing hard screening corrections 
to the hard amplitude for double--gap diffractive production.
Shown is only the generic structure of the partonic ladders;
the dominant contribution comes from gluons.}
\label{fig:hardscreen}
\end{figure}

One specific mechanism which can lead to qualitative changes of the
hard scattering process at high energies is ``local'' absorption of 
the hard scattering amplitude for diffractive production. 
This corresponds to the familiar attenuation of the hard gluon exchange 
diagram for Higgs production, but by a two--gluon ladder (see 
Fig.~\ref{fig:hardscreen}). Here the two--gluon ladder is attached to the 
partons which emitted the gluons involved in the Higgs production. 
Actually, in order to regularize the infrared divergences present in the 
two--gluon ladder we may consider instead the amplitude for the scattering 
of two colorless dipoles, in which the typical virtuality of the 
``constituents'' is $k_\perp^2$; a second parton with such virtuality 
is anyway present as a result of QCD evolution.
The amplitude of the additional interaction is then suppressed 
relative to the original hard amplitude by a factor
$\alpha_s^2 (k_\perp^2 /\tilde k_\perp^2) (x_0/x)^{\lambda}$, 
where $\tilde k_\perp$ is the characteristic gluon transverse 
momentum in the absorptive ladder. The characteristic
$\tilde k_\perp^2$ increases with energy, and $\lambda \approx 0.2 - 0.25$ at 
$Q^2 = 4\, \text{GeV}^2$ \cite{Ciafaloni:2003rd,Altarelli:2003hk}. 
Ultimately, the amplitude of the additional interaction would thus
reach a strength comparable to the maximal one (BDL), resulting in 
complete suppression of RGS. 

%
%
A semi-realistic estimate, using the eikonal 
approximation for the scattering of the colorless dipoles and 
assuming $\lambda \approx 0.2-0.25$, suggests that at the LHC energy 
the additional suppression is significant (a factor 2--3) 
for Higgs boson production by valence gluons, but much less so
for processes initiated by valence quarks; at the Tevatron energy 
the effect should be much less pronounced for both gluons and quarks. 
In order to evaluate this additional suppression at LHC energies 
quantitatively, one needs to follow the space--time evolution of the 
production of small--$x$, large--virtuality partons, and must not restrict 
the discussion to the gluon GPDs in the individual protons.

We emphasize that the screening effect described here is not included 
in the definition of the RGS probability due to soft interactions,
as it corresponds to a modification of the probability of
finding two gluons at small transverse distances without
reference to the spectator interactions. Detailed calculations 
of the RGS probability including this effect require modeling 
of the color connections in the parton wavefunctions of the protons.
Another perturbative screening correction to the hard process,
due to diffraction into high--mass states (with the mass increasing
with the collision energy) was considered in Ref.~\cite{Bartels:2006ea}, 
which leads to a complementary suppression of the amplitude for exclusive 
double--gap diffraction.
\subsection{Correlations between hard and soft interactions}
\label{subsec:correlations}
In our treatment of RGS in Section~\ref{subsec:combining} we have not 
taken into account effects due to correlations between partons in the 
wavefunctions of the colliding protons. Generally speaking, in the 
presence of such correlations the selection of particular configurations 
by the hard scattering process changes soft interactions as compared to 
``average'' configurations. The neglect of this change is basic
to the approximation (\ref{hard_soft_commute}), in which hard and 
soft interactions are assumed to commute also with respect to transverse
degrees of freedom. A consistent treatment of correlation effects requires 
a much more detailed description of hard and soft interactions and the
parton--hadron interface than the one given in Section~\ref{sec:gap_survival}.
Here we would like to discuss this problem at the qualitative level,
preparing the ground for a future in-depth investigation.

As a pedagogical example illustrating the effect of correlations
on the RGS probability, let us consider double--gap diffractive 
proton--deuteron ($pd$) scattering --- possibly including the quasi--elastic 
channel --- within the framework of Eq.~(\ref{survb}). 
The profile function for $pd$ elastic
scattering is significantly smaller than that for $pp$ scattering
even for energies at which $pp$ scattering is close to the black--disk
limit, 
\beq
\Gamma^{dp} \;\; \ll \;\; \Gamma^{pp} \;\; \lesssim \;\; 1. 
\label{Gamma_deuteron}
\eeq
This follows from Eq.~(\ref{unitarity}) when noting that the $dp$ 
total cross section is approximately equal to twice the $pp$ cross section, 
while the spatial size of the deuteron is much larger than the proton 
radius, and can be understood simply 
as the result of a ``dilution'' of the blackness of the individual 
protons due to their transverse motion in the deuteron bound state.
At the same time, the transverse spatial distribution of gluons is
now characterized by the transverse size of the deuteron. 
Because of Eq.~(\ref{Gamma_deuteron}), the factor $|1 - \Gamma^{dp}|^2$
in Eq.~(\ref{survb}) is always of order unity, and no significant
suppression takes place as it does in $pp$ scattering. Thus, one would
conclude that the RGS probability is much larger in $dp$ than in $pp$
scattering, which is clearly nonsensical. The paradox is resolved
by noting that hard and soft interactions in diffractive $pd$ scattering 
are highly correlated. By considering events with a hard process 
one is effectively selecting configurations
in which the projectile proton scatters from one of the nucleons.
Soft interactions in these configurations are significantly
larger than in $pd$ scattering in average configurations, in which
there is a substantial chance of the projectile ``missing'' the 
nucleons in the deuteron \footnote{It is worth noting here that diffraction 
in this case does not result from the presence of the fluctuations of the 
cross section strength. As a result, the cross section of diffraction is zero
at $t = 0$. At the same time, the diffractive cross section 
($dp\rightarrow pnp$) is not suppressed for $|t| \ge 1/R_d^2$, 
illustrating that diffraction cannot be described within the logic 
of the eigenstate scattering formalism \cite{Good:1960ba,Miettinen:1978jb} 
for finite $t$.  Also, there is no simple relation in this case between 
the Fourier transform of the diffractive amplitude (impact parameter 
representation) and the actual values of $b$ in the process;
the Fourier transform would suggest 
$\langle b^2 \rangle \propto 1/B$ [$B$ is the slope of the $pp$
elastic cross section, Eq.~(\ref{B_zero})], while in reality
$\langle b^2 \rangle \propto R_d^2$.}.

The deuteron example shows that transverse correlations 
in the wavefunctions can qualitatively change the picture of RGS.
In particular, positive spatial correlations between hard partons 
and the opacity for soft interactions decrease the RGS probability 
compared to the uncorrelated estimate based on Eq.~(\ref{survb}).
In connection with $pp$ scattering, Eq.~(\ref{exclusive_diffraction}),
we now discuss the effects of two types of transverse correlations:
(a) long--distance correlations due to scattering from the 
proton's long--range pion field (``pion cloud''), (b) short--distance 
correlations related to parton clustering in ``constituent quarks''.

A distinctive contribution to diffractive $pp$ scattering at small
$\bm{p}_{1\perp}'$ results from the process in which a soft pion,
emitted and absorbed by proton 1, scatters diffractively from proton 2.
This contribution could properly be calculated using the known pion--nucleon 
coupling, and applying Eqs.~(\ref{amp_momentum})--(\ref{amp_momentum_t}) 
to the $\pi p$ diffractive amplitude, with the two--gluon 
formfactor in the pion and the profile function of $\pi p$ elastic scattering.
It is generally small, for two reasons. First, the coupling 
of the soft pion to the nucleon is small because it is the Goldstone boson
of spontaneously broken chiral symmetry. Second, softness of the pion implies 
that its longitudinal momentum 
fraction in the proton be small, $y \lesssim m_\pi / m_p$. In Higgs
production at the LHC, where $x_{1, 2} \sim 10^{-2}$, this puts the 
momentum fraction of the gluons in the pion at relatively large values,
$z \sim x_{1, 2}/y \sim 10^{-1}$, where the gluon distribution 
is not enhanced by DGLAP evolution. 

In the partonic picture, the pion cloud contribution represents
the result of specific correlations between hard and 
soft partons in the proton wavefunction. This observation has an
interesting consequence for the estimate of the RGS probability
for $pp$ diffractive scattering based on Eq.~(\ref{survb}).
Namely, the gluon distribution in the proton receives a 
contribution from the pion cloud at transverse distances 
$\rho \sim 1/(2 m_\pi)$ and momentum fractions 
$x \lesssim m_\pi / m_N$ \cite{Strikman:2003gz}. Including this
contribution in Eq.~(\ref{survb}) would be inconsistent, 
since a hard process involving these gluons in the pion cloud
should be accompanied by a very specific modification of the 
soft interactions, which is not accounted for in Eq.~(\ref{survb}).
For this reason, we have not included an explicit pion cloud 
contribution in the two--gluon formfactor parametrization used
in our estimate leading to the value (\ref{S2_LHC_num}). 
More precisely, if we knew the ``physical'' gluon GPD,
which by definition includes the pion cloud contribution, 
we would need to remove this contribution before using 
the GPD in Eq.~(\ref{survb}), and thus obtain a lower value
for the RGS probability. This is just another example of the
general rule that correlations lower the 
RGS probability compared to the independent interaction
approximation, Eq.~(\ref{survb}).

%
%
\begin{figure}
\includegraphics[width=8cm]{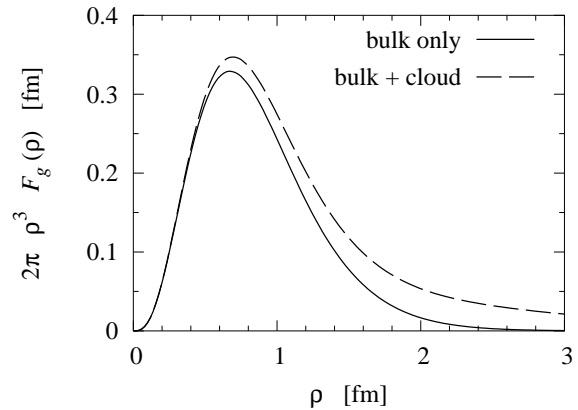}
\caption[]{The two--component parametrization of the 
transverse spatial distribution of gluons including the pion cloud,
Eqs.~(\ref{fg_twocomp}) and (\ref{fg_cloud}). Shown are the radial 
distributions $2\pi \rho^3 F_g (\rho )$, 
the integral of which determines the average gluonic transverse 
size, $\langle \rho^2 \rangle$. For the sake of comparison, 
the figure shows the combined (bulk + cloud) distribution
without adjustment of the normalization factor ($N = 1$),
so that the bulk contribution is the same as in the 
case of no pion cloud.}
\label{fig:twocomp_fg_rho2}
\end{figure}
\begin{figure}
\includegraphics[width=8cm]{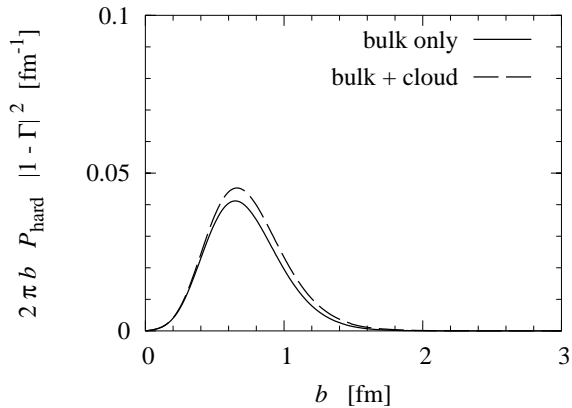}
\caption[]{The integrand (impact parameter distribution) 
in the rapidity gap survival probability, Eq.~(\ref{survb}), 
for the two--component form of the transverse spatial distribution
of gluons. The parameters are the same as in Fig.~\ref{fig:twocomp_fg_rho2}.} 
\label{fig:twocomp_survb}
\end{figure}

To estimate the correction resulting from the removal of the 
pion cloud contribution from the gluon GPD, we evaluate
Eq.~(\ref{survb}) with a ``two--component'' parametrization 
of the two--gluon formfactor (\textit{viz}.\ the transverse spatial
distribution of gluons) including the pion cloud. 
We write
\beq
F_g (\rho ) \;\; = \;\; N \left[ F_{\text{bulk}} (\rho ) \; + \; 
F_{\text{cloud}} (\rho ) \right] ,
\label{fg_twocomp}
\eeq
where $N$ is a factor which ensures overall normalization
to $\int d^2\rho F_g(\rho) = 1$. The ``bulk'' contribution to the 
two--gluon formfactor we parametrize by the dipole formfactor with 
$m_g^2 = 1 \, \text{GeV}^2$, Eq.~(\ref{twogl_dipole}),
the  ``cloud'' contribution as
\beq
F_{\text{cloud}} (\rho ) \;\; = \;\;
C_{\text{cloud}} \; \frac{\rho^2}{\rho^2 + \rho_0^2} \;
\frac{e^{-2 m_\pi \rho}}{2 m_\pi \rho} .
\label{fg_cloud}
\eeq
This form is essentially the asymptotic form of the gluon
density at $\rho \gtrsim 1/(2 m_\pi)$ for $x \ll m_\pi / m_N$  
(``Yukawa tail'') \cite{Strikman:2003gz}, regularized at 
small $\rho$ such as to avoid a large contribution in the
bulk region; the parameter $\rho_0^2$ is chosen of the
order $\langle \rho^2 \rangle_{g, \text{bulk}} 
= 8/m_g^2 = 0.3\, \text{fm}^2$. The coefficient $C_{\text{cloud}}$
we determine such that the inclusion of the cloud contribution
increases the overall gluonic transverse size of the proton by 30\% ,
which is the value found in the calculation of 
Ref.~\cite{Strikman:2003gz} (based on the phenomenological 
gluon distribution in the pion) and supported by the 
$J/\psi$ photoproduction data; see Section~\ref{sec:transverse_hard}.
Figure~\ref{fig:twocomp_fg_rho2} shows the transverse spatial distributions
for ``bulk only'' and ``bulk + cloud'', multiplied by 
$2\pi \rho^3$, the integral of which determines 
$\langle \rho^2 \rangle_g$. Figure~\ref{fig:twocomp_survb} shows 
the impact parameter distributions in the RGS probability
(\textit{cf.}\ Fig.~\ref{fig:survb}) obtained in the two cases. 
One sees that removal of the pion cloud contribution indeed reduces
the RGS probability. The numerical effect turns out to be rather 
small,
\beq
\frac{S^2 \text{(bulk only)}}{S^2 \text{(bulk + cloud)}}
\;\; = \;\; 0.94 .
\eeq
This can be explained by the fact that the pion cloud contribution
to the gluon density is noticeable compared to the bulk only at 
transverse distances $\rho \gg 1/(2 m_\pi) = 0.7 \, \text{fm}$,
while the RGS integral is dominated by rather short distances,
$\rho_{\text{eff}} \sim b_{\text{eff}}/2 \sim 0.4 \, \text{fm}$,
\textit{cf}\ Figs.~\ref{fig:coord} and \ref{fig:survb}.
%
%

Corrections to the rapidity gap survival probability as given by 
Eq.~(\ref{survb}) can also come from transverse short--distance 
correlations in the proton wavefunction (correlation length $\ll$
proton size). The Tevatron CDF data on $\bar pp$ collisions with 
multiple hard processes (three jet plus photon production) \cite{Abe:1997bp} 
indicate the presence of significant correlations between the 
transverse positions of hard partons over distances 
$r \sim 0.3\, \text{fm}$ \cite{Frankfurt:2004kn}. 
Given one hard parton with $x \gtrsim 0.05$ at a certain transverse position, 
it is much more likely to find in the proton wavefunction 
a second hard parton $x \gtrsim 0.05$ within a distance $\sim r$ 
than elsewhere in the transverse plane. As a result, in $pp$ events with 
(at least) one hard process the probability for a second hard interaction is 
substantially larger (by a factor of $\sim 2$) than it would be without
correlations. One may also suppose that the local cross section density
(opacity) for soft inelastic interactions is higher near 
the position of a hard parton than elsewhere.
Such correlations would result in a higher probability for
inelastic interactions in $pp$ events with a hard process
(such as the hard two--gluon exchange in double--gap diffractive
production) as compared to generic $pp$ collisions, and would 
thus decrease the RGS probability compared to Eq.~(\ref{survb}).

The corrections to Eq.~(\ref{survb}) due to short--distance correlations
can be viewed as an effective reduction of the size of the diffractively
scattering system, from the proton radius to the size of the transverse
correlation, $r$. The corrections could be particularly large in
the situation when the correlated areas are ``black spots,'' while the
proton overall is still ``gray'' because of the dilution by
the transverse motion. This situation would in a sense correspond to 
the above example of the deuteron,
with the proton now playing the role of the deuteron.
A quantitative description of these effects would require
detailed modeling of the correlation between hard partons 
and the opacity for soft inelastic interactions, including an
analysis of $pp$ elastic scattering allowing local fluctuations 
in opacity, which are outside of the scope of the present paper.

A model of the proton accounting for short--distance correlations 
between partons is the so--called chiral quark--soliton model
\cite{Diakonov:1987ty}, which describes the proton as a system of 
constituent quarks bound by a classical pion field, 
see Ref.~\cite{Diakonov:2002fq} for a review. This model implements 
the short--distance scale related to the spontaneous breaking of 
chiral symmetry, which appears here as the ``size'' of the constituent 
quark, $r \sim 0.3\, \text{fm}$ (in the Euclidean formulation of QCD
this scale can be associated with the average instanton size in the vacuum).
This model provides a consistent description of the 
twist--2 quark-- and antiquark distributions in the nucleon
at the scale $\mu \sim r^{-1} \approx 600\, \text{MeV}$ \cite{Diakonov:1996sr}.
It also suggests that the gluons at this scale are ``packaged'' inside the 
constituent quarks and antiquarks \cite{Diakonov:2002fq}. Assuming 
perturbative QCD evolution to be applicable starting from this scale 
one would thus expect significant correlations between the positions
of hard quarks and gluons and the opacity for soft interactions
over a transverse size $r \sim 0.3 \, \text{fm}$. 
Incorporating these correlations
into the description of the BDL in high--energy $pp$ scattering 
and the theory of RGS in diffractive processes is an important
problem for future studies.

To summarize, the dynamical mechanisms which we have discussed here 
(and indeed all mechanism which we are aware of) give rise to 
positive correlations between the transverse position of hard partons 
and the opacity for inelastic interactions. We can thus say with some 
confidence that estimates based on the independent interaction 
approximation, Eq.~(\ref{survb}), represent an upper bound 
on the RGS probability.
\section{Transverse momentum dependence}
\label{sec:differential}
The interplay of hard and soft interactions in exclusive double--gap
diffraction not only causes the suppression of the integrated
cross section summarized in the RGS probability, but also 
gives rise to a distinctive dependence of the cross section
on the final proton transverse momenta. By observing this 
``diffraction pattern'' one can perform detailed tests of the 
diffractive reaction mechanism, and even extract information
about the two--gluon formfactors of the colliding protons.
The transverse momentum dependence also contains information
about the quantum numbers (parity) of the produced system 
\cite{Kaidalov:2003fw}. We consider here production of a $0^+$ system,
for which the hard scattering amplitude depends on the proton transverse 
momenta only through $|\bm{p}_{1\perp}'|$ and $|\bm{p}_{2\perp}'|$, 
and the diffractive amplitude is given by Eq.~(\ref{amp_momentum});
we comment on production of $0^-$ systems at the end of this section.

%
%
\begin{figure*}
\includegraphics[width=16.5cm]{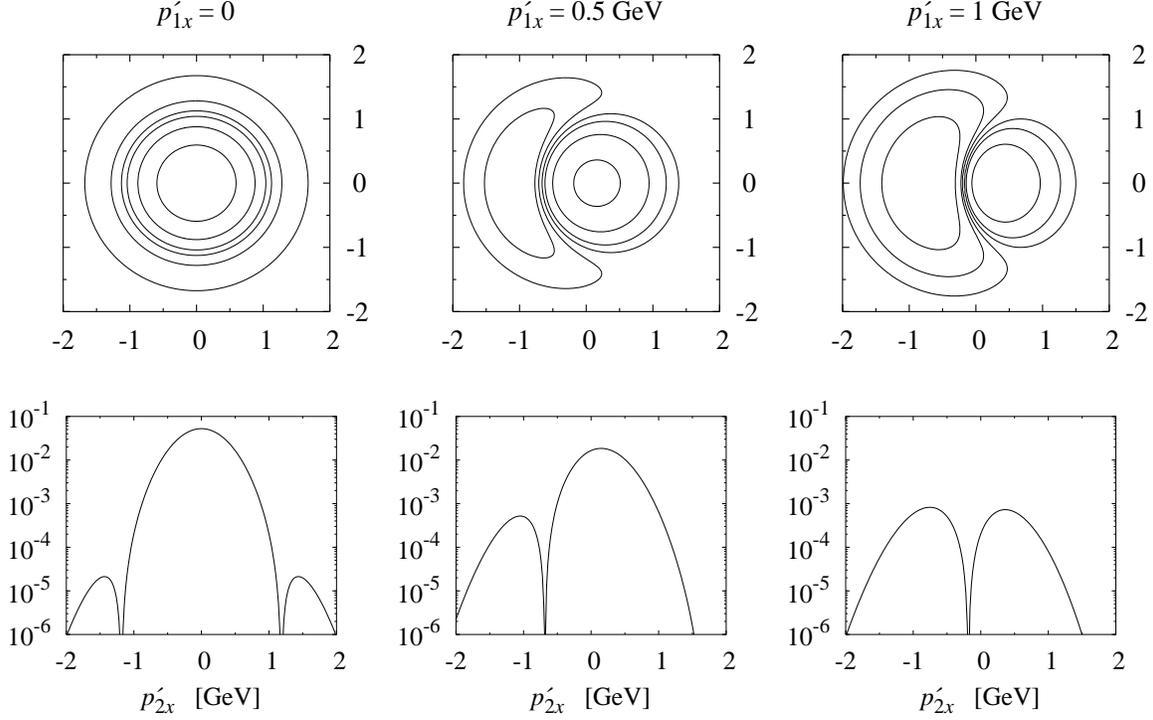}
\caption[]{Transverse momentum dependence of the cross section
for double--gap exclusive diffraction (\ref{exclusive_diffraction}).
The plots show the modulus squared of the amplitude, $|T_\text{diff}|^2$,
with $\kappa = 1$, as a function of $\bm{p}_{2\perp}'$, 
for three fixed values of $\bm{p}_{1\perp}'$ (chosen to point in 
$x$--direction, $p_{1y}' = 0$). The upper row of plots
show the contours of constant values of $|T_\text{diff}|^2$
on a logarithmic scale as functions of $p_{2x}'$ and $p_{2y}'$ 
(in units of GeV). The lower row of plots shows the profile
along the $p_{2y}' = 0$ axis. Shown are the results calculated
with the exponential parametrizations of the two--gluon formfactor
and the $pp$ elastic amplitude, Eq.~(\ref{T_diff_gauss_p1_p2}).
The diffraction pattern in $\bm{p}_{2\perp}'$ evolves from a 
rotationally symmetric one for $\bm{p}_{1\perp}' = 0$ to a 
two--centered one for large $|\bm{p}_{1\perp}'|$.}
\label{fig:diffp}
\end{figure*}
For a quick orientation over the transverse momentum dependence,
we evaluate the amplitude
with the exponential parametrizations of the two--gluon formfactor,
Eqs.~(\ref{twogl_exponential}), and the $pp$ elastic 
scattering amplitude, Eq.~(\ref{Gamma_gaussian}).
In this case the convolution integral
in Eq.~(\ref{amp_momentum}) reduces to a Gaussian integral,
and we obtain a closed expression for the amplitude,
\be
\lefteqn{T_{\text{diff}} (\bm{p}_{1\perp}', \bm{p}_{2\perp}') } &&
\nonumber \\[1ex]
&=& 
\kappa \; \exp \left( - \frac{B_{g1} \bm{p}_{1\perp}^{'2}}{2} 
- \frac{B_{g2} \bm{p}_{2\perp}^{'2}}{2} \right) 
\nonumber \\
&\times &
\left\{ 1 - \frac{B}{B_{\text{tot}}}
\exp \left[ \frac{(B_{g1} \, \bm{p}_{1\perp}' 
- B_{g2} \, \bm{p}_{2\perp}')^2}
{2 B_{\text{tot}}} \right]
\right\} ,
\label{T_diff_gauss_p1_p2}
\ee
where $B_{g1} \equiv B_g (x_1)$ and $B_{g2} \equiv B_g (x_2)$ 
are the slopes corresponding to the momentum fractions 
$x_{1, 2} \sim \xi_{1, 2}$, and
\beq
B_{\text{tot}} \;\; \equiv \;\; B_{g1} + B_{g2} + B .
\eeq
[As a check, we note that the integral of the square of 
this expression, divided by the corresponding expression for 
$B = 0$ (no soft interactions), reproduces the result for the RGS 
probability obtained in the coordinate space calculation, 
Eq.~(\ref{S2_exponential_gen}).] The amplitude (\ref{T_diff_gauss_p1_p2}) 
vanishes trivially for large $|\bm{p}_{1\perp}|$ or $|\bm{p}_{2\perp}|$, 
independently of the directions of the momentum vectors. 
In addition, it has a zero at finite values of the transverse
momenta, namely at
\be
\lefteqn{(B_{g1} \, \bm{p}_{1\perp}' - B_{g2} \, \bm{p}_{2\perp}')^2 }
&& 
\nonumber \\
&=& 2 B_{\text{tot}}
\; \ln \frac{B_{\text{tot}}}{B} 
\nonumber \\
&\approx & 2 (B_{g1} + B_{g2}) 
\hspace{3em} (B \gg B_{g1}, B_{g2}) .
\label{zero}
\ee
This zero arises because of the destructive interference of
the amplitude of the hard scattering process alone and the 
amplitude including soft elastic rescattering, 
\textit{cf.}\ Fig.~\ref{fig:mom}, and directly reflects
the interplay of hard and soft interactions. It leads to a dip
in the diffractive cross section, and thus to a typical 
``diffraction pattern'' in the dependence on the transverse momentum
of the first proton, $\bm{p}_{1\perp}'$, at fixed transverse momentum
of the second proton, $\bm{p}_{2\perp}'$. Figure~\ref{fig:diffp}
shows this diffraction pattern in the kinematics of Higgs production 
at the LHC at zero rapidity, for which $B_{g1} = B_{g2}$. 
One sees that the pattern 
in $\bm{p}_{2\perp}'$ evolves from a rotationally symmetric one for 
$\bm{p}_{1\perp}' = 0$ to a two--centered one for large $|\bm{p}_{1\perp}'|$.
This basic feature of the diffractive cross section does not depend on the
details of the parametrization of the two--gluon formfactor; similar
results are obtained with the dipole parametrization,
\textit{cf.}\ the detailed comparison below.

%
%
\begin{figure}
\includegraphics[width=8cm]{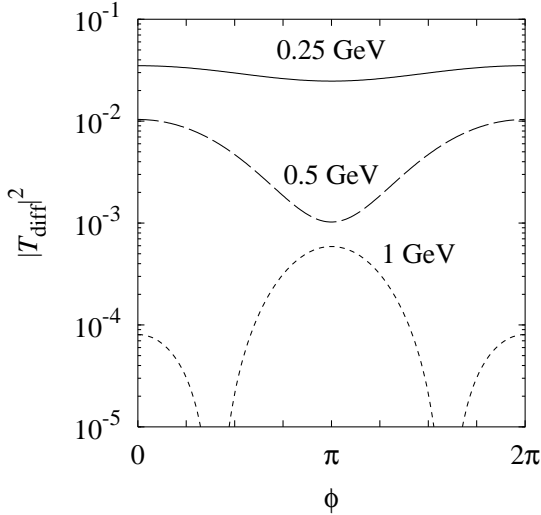}
\caption[]{Angular dependence of the cross section
for double--gap diffractive production of a $0^+$ system.
The plot shows the modulus squared of the amplitude,
$|T_{\text{diff}}|^2$, Eq.~(\ref{amp_momentum}), with $\kappa = 1$. 
The numbers above the curves indicate the values of
$|\bm{p}_{1\perp}'| = |\bm{p}_{1\perp}'|$. The kinematics 
corresponds to Higgs boson production at the LHC.}
\label{fig:phidep}
\end{figure}
The diffraction pattern of Fig.~\ref{fig:diffp} implies a 
strong angular dependence of the cross section, which, moreover,
changes with the magnitude of the transverse momenta, 
$|\bm{p}_{1\perp}'|$ and $|\bm{p}_{2\perp}'|$. This is illustrated
in Figure~\ref{fig:phidep}, which shows the dependence 
of $|T_{\text{diff}}|^2$ on the
angle between the transverse  momenta, $\phi$, for various 
values of $|\bm{p}_{1\perp}'| = |\bm{p}_{2\perp}'|$.
For small values of the transverse momenta the cross section 
is maximal at zero angle; for large values (where the cross sections
as a function of angle runs through the diffractive dip) it is maximal
at $\phi = \pi$. This dependence needs to be taken into account when
attempting to maximize the diffractive cross section in the search
for new particles.

%
%
\begin{figure*}
\begin{tabular}{cc}
\includegraphics[width=7cm]{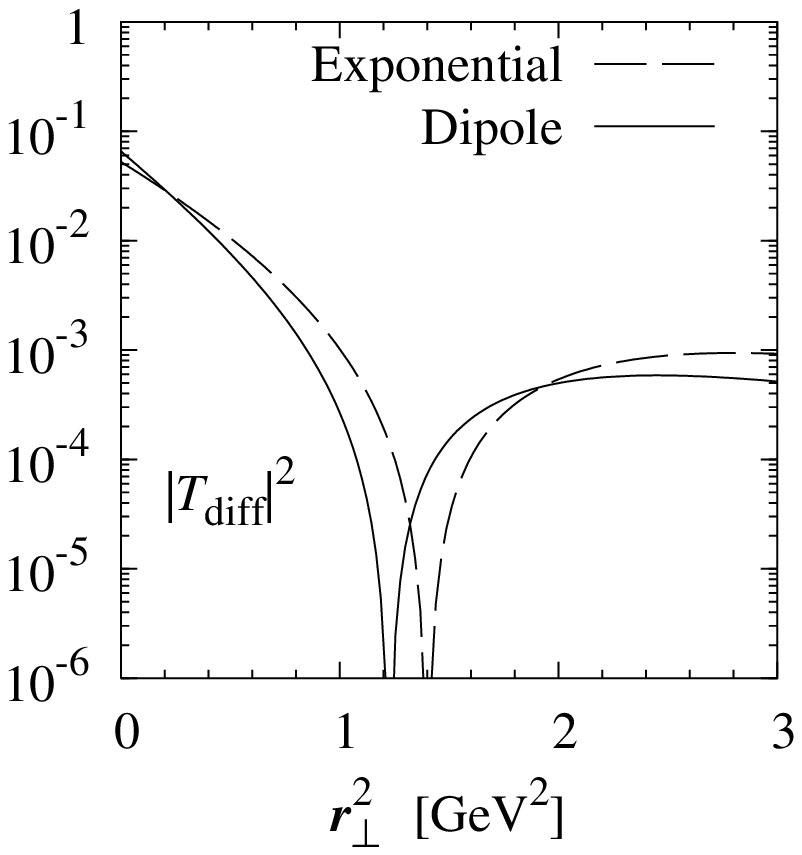}
&
\includegraphics[width=7cm]{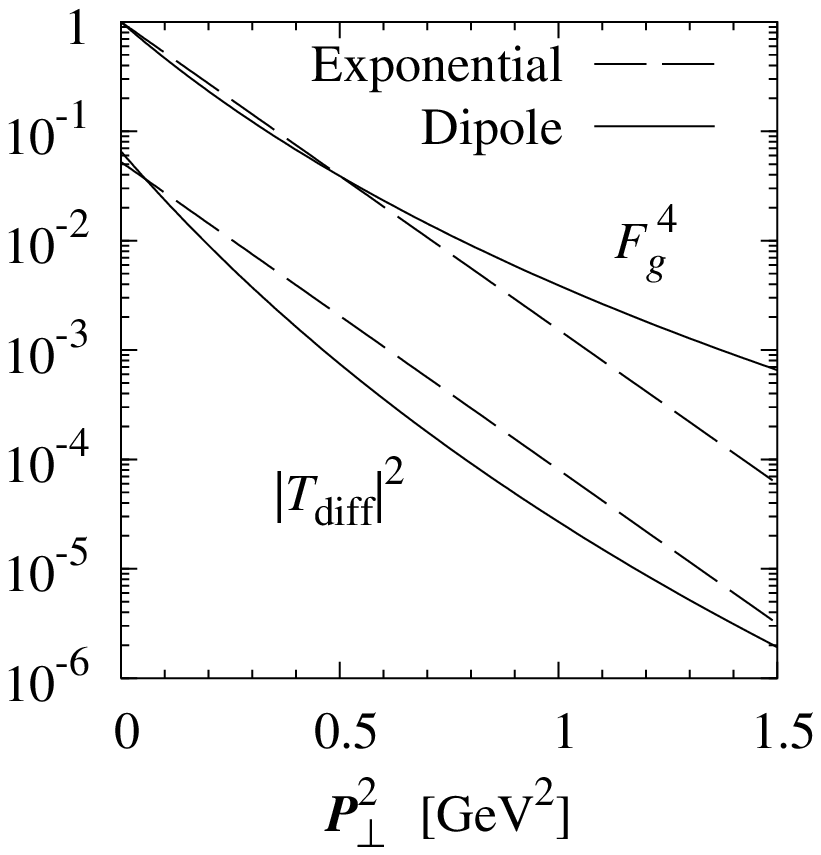}
\end{tabular}
\caption[]{\textit{Left:}
Dependence of the cross section for double--gap diffractive
$0^+$ production on the square of the difference of the proton transverse 
momenta, $\bm{r}_\perp^2$, for $\bm{P}_\perp = 0$ (\textit{i.e.},
$\bm{p}_{1\perp}' = -\bm{p}_{2\perp}' = 2\bm{r}_\perp$).
The kinematics corresponds to production of a system with 
$M = 140\, \text{GeV}$ at the LHC energy at zero rapidity 
($x_1 = x_2 \sim 10^{-2}$).
The plot shows the modulus squared of the amplitude,
$|T_{\text{diff}}|^2$, Eq.~(\ref{amp_momentum}), with $\kappa = 1$,
obtained with the exponential and dipole parametrizations of
the two--gluon formfactor. For the exponential parametrization, 
the position of the
diffractive zero is given by Eq.~(\ref{zero}).
\textit{Right:} Dependence on the square of the sum of the 
transverse momenta, $\bm{P}_\perp^2$, for $\bm{r}_\perp = 0$ (\textit{i.e.},
$\bm{p}_{1\perp}' = \bm{p}_{2\perp}' = \bm{P}_\perp$).
Also shown is $F_g^4 (t)$ for the
two parametrizations, which would be the $t$--dependence of the 
amplitude without soft spectator interactions.}
\label{fig:rpdep}
\end{figure*}
An interesting questions is which specific features of the transverse 
momentum dependence of the diffractive cross section could be used to 
test the diffractive reaction mechanism, and possibly extract information
about the two--gluon formfactors of the colliding protons. Such studies 
would be feasible in diffractive dijet production, which has a relatively 
large cross section and also allows one to vary the invariant mass of the 
diffractively produced system, and thus the effective values of the gluon 
momentum fraction, $x_1$ and $x_2$. To address this question,
we again start from the explicit expression for the amplitude obtained
with the exponential parametrization of the two--gluon formfactor,
Eq.~(\ref{T_diff_gauss_p1_p2}). We rewrite it in terms of
the center--of--mass and relative transverse momenta of the
final--state protons,
\be
\bm{P}_\perp &\equiv& (\bm{p}_{1\perp}' + \bm{p}_{2\perp}')/2, 
\nonumber \\
\bm{r}_\perp &\equiv& \bm{p}_{1\perp}' - \bm{p}_{2\perp}' ,
\ee
and obtain
\be
\lefteqn{T_{\text{diff}} (\bm{p}_{1\perp}', \bm{p}_{2\perp}') } &&
\nonumber \\[1ex]
&=& 
\kappa \; \exp \left[ - \frac{B_{g1} + B_{g2}}{2} \left( 
\bm{P}_\perp^2 + \frac{\bm{r}_\perp^2}{4} \right) 
\right. \nonumber \\
&& \left. 
\phantom{\exp} \;
- \frac{B_{g1} - B_{g2}}{2} (\bm{P}_\perp \bm{r}_\perp) \right]
\nonumber \\
&\times &
\left\{ 1 - \frac{B}{B_{\text{tot}}}
\exp \left[ 
\frac{(B_{g1} - B_{g2})^2}{2 B_{\text{tot}}} \bm{P}_\perp^2 
\right.\right. 
\nonumber \\
&& + \frac{(B_{g1} + B_{g2})^2}{2 B_{\text{tot}}} \frac{\bm{r}_\perp^2}{4} 
\nonumber \\
&& \left. \left.
+ \frac{(B_{g1} - B_{g2}) (B_{g1} + B_{g2})}{2 B_{\text{tot}}} 
(\bm{P}_\perp \bm{r}_\perp) \right]
\right\} .
\label{T_diff_P_r}
\ee
For production at zero rapidity, for which $x_1 = x_2 \equiv x$,
and $B_{g1} = B_{g2} \equiv B_g$, this simplifies to
\be
\lefteqn{T_{\text{diff}} (\bm{p}_{1\perp}', \bm{p}_{2\perp}') } &&
\nonumber \\[1ex]
&=& 
\exp \left( -B_g \bm{P}_\perp^2 - \frac{B_g \bm{r}_\perp^2}{4} \right) 
\nonumber \\
&\times &
\left[ 1 - \frac{B}{B_{\text{tot}}}
\exp \left( 
\frac{B_g^2}{B_{\text{tot}}} \bm{r}_\perp^2 \right) \right] ,
\label{T_diff_P_r_sym}
\ee
where now $B_{\text{tot}} = B + 2 B_g$. In this case the amplitude
does not depend on the variable $(\bm{P}_\perp \bm{r}_\perp)
= (\bm{p}_{1\perp}^{'2} - \bm{p}_{2\perp}^{'2})/2$, which is 
an obvious consequence of its symmetry with respect to the 
interchange of $\bm{p}_{1\perp}^{'}$ and $\bm{p}_{2\perp}^{'}$.
Furthermore, one sees that the dependences of Eq.~(\ref{T_diff_P_r_sym})
on $\bm{P}_\perp^2$ and $\bm{r}_\perp^2$ are very different. 
The dependence on $\bm{P}_\perp^2$ is monotonic and governed by 
the parameter $B_g$ alone; it essentially
probes the square of the two--gluon formfactors of the colliding protons.
The dependence on $\bm{r}_\perp^2$, however, is governed by both
$B_g$ and $B$, and exhibits the diffractive zero (\ref{zero});
it reflects the interplay of hard and soft interactions.

The qualitative differences between the $\bm{P}_\perp$ and $\bm{r}_\perp$ 
dependence of the amplitude are not specific to the exponential
parametrization of the two--gluon formfactor, and can be used
to test the diffractive reaction mechanism and extract information 
about the two--gluon formfactor. Figure~\ref{fig:rpdep} (left plot) 
shows the squared modulus of the amplitude (for $\kappa = 1$) 
as a function of $\bm{r}_\perp^2$,
for $\bm{P}_\perp = 0$, in the kinematics corresponding to 
dijet production with $x = 10^{-2}$ at the LHC. One sees that the
dependence in the forward peak (near $\bm{r}_\perp^2 = 0$), and even 
the position of the diffractive dip, are not very different for the
two parametrizations. Experimental observation of this structure would 
thus constitute a stringent test of the diffractive reaction mechanism.

Figure~\ref{fig:rpdep} (right plot) 
shows the dependence of the cross section for 
double--gap diffractive $0^+$ production on $\bm{P}_\perp^2$, for 
$\bm{r}_\perp = 0$. Also shown is the fourth power of the 
two--gluon formfactor, $F_g^4 (t)$ for the two parametrizations, 
which would be the $t$--dependence of the amplitude without soft 
spectator interactions. For the exponential parametrization the
$t$--dependence of the full diffractive amplitude is identical 
to that of $F_g^4 (t)$, \textit{cf.}\ Eqs.~(\ref{T_diff_P_r_sym})
and (\ref{twogl_exponential}); one sees that the two dependences
are similar also for the dipole parametrization. The different 
normalization of the two sets of curves reflects the RGS probability,
\textit{cf.}\ Sec.~\ref{subsec:geometry}.

The position of the diffractive zero in the $\bm{r}_\perp^2$--dependence
(see Fig.~\ref{fig:rpdep}, left plot) is correlated with the slope of the 
monotonic $\bm{P}_\perp^2$--dependence of the diffractive cross section;
both essentially reflect the two--gluon formfactor of the colliding
protons. A sensible strategy for the analysis of double--gap 
diffractive dijet 
production would be to first establish the existence of the 
diffractive zero in $\bm{r}_\perp^2$ at $\bm{P}_\perp = 0$, 
and then extract $B_g$ from the $\bm{P}_\perp^2$--dependence of the 
diffractive cross section at $\bm{r}_\perp = 0$, where the
cross section is maximal. In the next step, one could change the
dijet mass (\textit{i.e.}, the momentum fractions $x_1 = x_2 = x$
in the two--gluon formfactor) and verify whether both the position 
of the zero and the $\bm{P}_\perp^2$--slope change proportionately,
and whether the rate of change with $\ln x$ is consistent with
the value of $\alpha'_g$ at the relevant scale, \textit{cf.}\ 
Eq.~(\ref{alpha_g}).

The $x_{1, 2}$--dependence of the two--gluon formfactor in the protons
can be probed more directly by extending the measurements of diffractive 
production to non-zero rapidity, $y \neq 0$, corresponding to different 
momentum fractions of the annihilating gluons in the two protons,
\beq
x_{1, 2} \;\; = \;\; x \, e^{\pm y} , 
\hspace{3em} x \;\; \equiv \;\; \frac{m_H}{\sqrt{s}}.
\label{x_1_2_from_y}
\eeq
In this case the two--gluon formfactors of the two protons are
different, and the diffractive cross section is no longer 
invariant under exchange of the final proton transverse momenta,
$\bm{p}_{1\perp}' \leftrightarrow \bm{p}_{2\perp}'$. One sees that
the expression for the amplitude obtained with the exponential 
parametrization, Eq.~(\ref{T_diff_P_r_sym}), acquires a
dependence on $(\bm{P}_\perp \bm{r}_\perp) = 
(\bm{p}_{1\perp}^{'2} - \bm{p}_{2\perp}^{\prime 2})/2$, which is controlled
by the difference of the slopes, $B_{g1} - B_{g2}$. With the
$x$--dependence of the individual slopes given by
\be
B_{g1} &\equiv& B_g (x_1) \;\; = \;\; 
B_g (x) + 2\alpha'_g \ln \frac{x_1}{x} , \\
B_{g2} &\equiv& B_g (x_2) \;\; = \;\; 
B_g (x) + 2\alpha'_g \ln \frac{x_2}{x} , 
\ee
their difference is directly proportional to the rapidity
[\textit{cf.}\ Eq.~(\ref{x_1_2_from_y})],
\beq
B_{g1} - B_{g2} \;\; = \;\; 2 \alpha'_g y .
\eeq
For small rapidities, $|y| \lesssim 1$, and because of the relatively
small value of $\alpha'_g$ this difference is substantially smaller 
than the central value, $B_g (x)$, and the dependence of the 
amplitude on $(\bm{P}_\perp \bm{r}_\perp)$
can be treated in first--order approximation.
This implies that the cross section depends practically
linearly on $(\bm{P}_\perp \bm{r}_\perp)$ for $|y| \lesssim 1$.

%
%
\begin{figure}
\begin{tabular}{c}
\includegraphics[width=8cm]{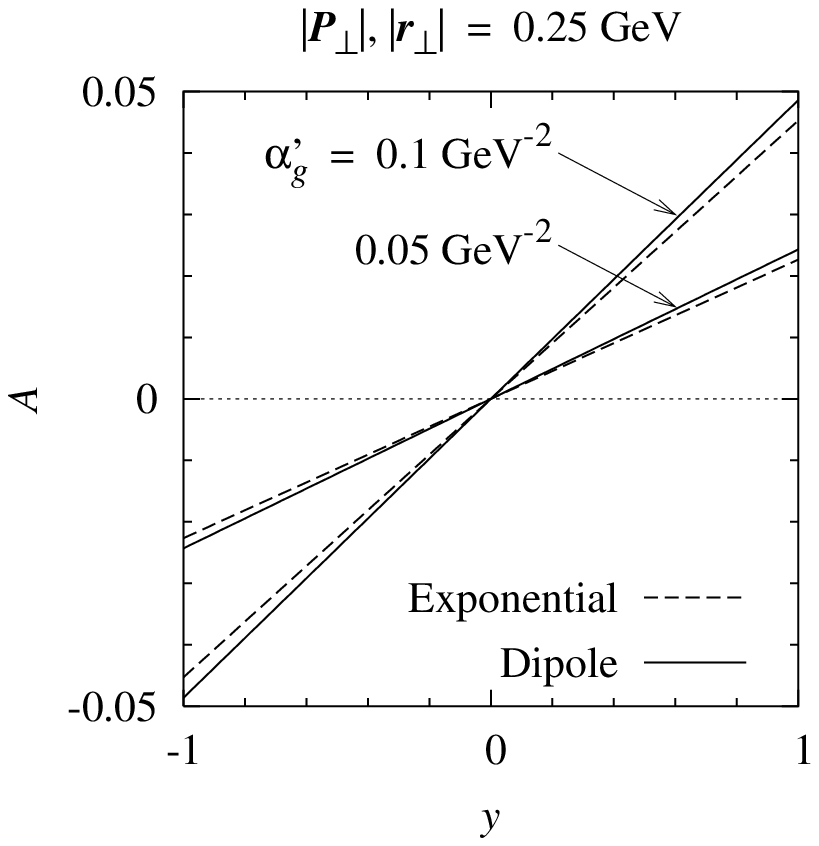} \\
\includegraphics[width=8cm]{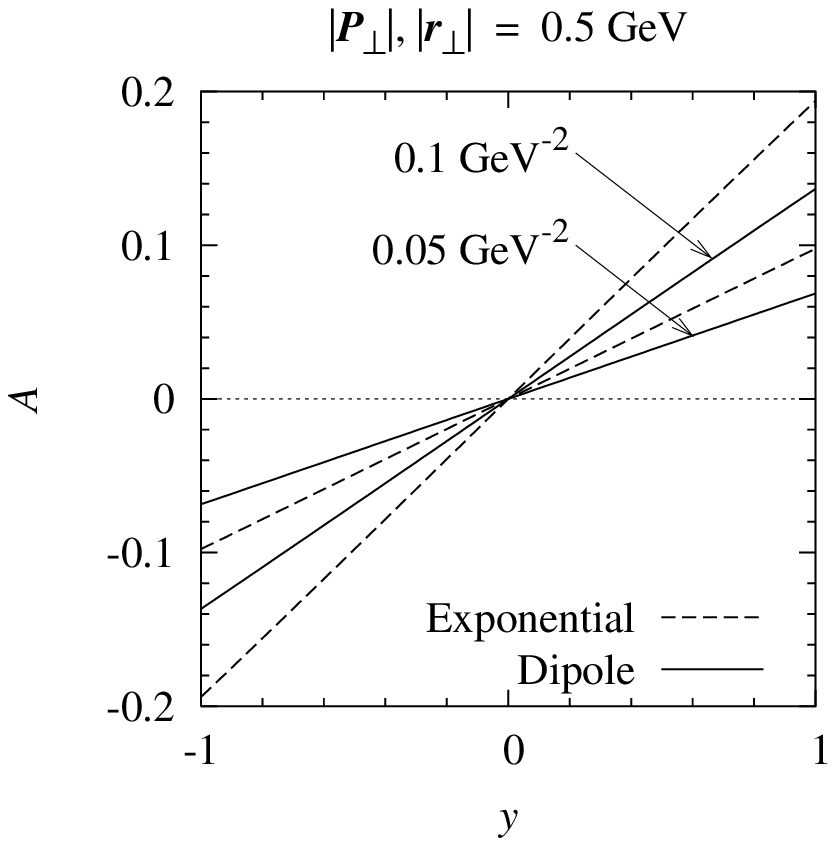}
\end{tabular}
\caption[]{The cross section asymmetry (\ref{asym}) as a function of
rapidity, $y$. Shown are the results obtained with the exponential and the 
dipole parametrizations of the two--gluon formfactor, 
for two representative values of $\alpha'_g$. In this calculation only 
the $y$--dependence of the cross section arising from the convolution 
integral is taken into account, not the $y$--dependence arising from the 
overall normalization, $\kappa$.}
\label{fig:asym}
\end{figure}
A convenient observable to measure is the asymmetry of the 
cross section with respect to $\bm{r}_\perp \rightarrow -\bm{r}_\perp$ 
and $\bm{P}_\perp \rightarrow \bm{P}_\perp$ (\textit{i.e.}, 
$\bm{p}_{1\perp}' \leftrightarrow \bm{p}_{2\perp}'$) at fixed
$x$ and $y \neq 0$ (\textit{i.e.}, fixed $x_1 \neq x_2$),
\be
A &\equiv& 
\frac{
\sigma_{\text{diff}}(\bm{p}_{1\perp}', \bm{p}_{2\perp}') -
\sigma_{\text{diff}}(\bm{p}_{2\perp}', \bm{p}_{1\perp}')}
{\sigma_{\text{diff}}(\bm{p}_{1\perp}', \bm{p}_{2\perp}') +
 \sigma_{\text{diff}}(\bm{p}_{2\perp}', \bm{p}_{1\perp}')} ;
\label{asym}
\ee
alternatively, one could exchange the rapidities and leave the
transverse momenta the same. This asymmetry is odd in $y$ and vanishes
linearly for $y \rightarrow 0$. For $|y| \lesssim 1$, it is practically
is proportional to $(\bm{P}_{\perp}\bm{r}_{\perp}) = 
(\bm{p}_{1\perp}^{\prime 2} - \bm{p}_{2\perp}^{\prime 2})/2$. 
When calculating the asymmetry at finite $y$, we have to take into
account that, in general, also the overall normalization of the cross 
section, $\kappa (s, \xi_1, \xi_2)$, changes with $y$, because of the
change of arguments in the $t = 0$ gluon GPDs of the protons, 
see Eq.~(\ref{kappa}). However, this change relative to the
value at $y = 0$ is of second order in $y$ (the changes in the 
arguments of the gluon densities cancel each other to first order)
and can be neglected for small $y$. This implies that for $y \ll 1$. 
the asymmetry is of the form
\be
A &\sim & C y \alpha'_g  (\bm{P}_{\perp}\bm{r}_{\perp}) ,
\ee
where the constant, $C$, is calculable solely from the convolution 
integral of the two--gluon formfactor and the $pp$ elastic amplitude,
and does not contain information on the gluon densities. For finite $y$,
the asymmetry is still proportional to 
$\alpha'_g (\bm{P}_{\perp}\bm{r}_{\perp})$, but the coefficient
is a more complicated function of $y$, which depends also on the
gluon densities in the colliding protons.
Figure~\ref{fig:asym} shows the theoretical
estimate of the asymmetry as obtained with the exponential and the dipole 
parametrizations of the two--gluon formfactor, as a function of the rapidity,
$y$, for two representative values of $\alpha'_g$. In this calculation,
for simplicity, we have taken into account only the $y$--dependence 
of the cross section arising from the RGS integral, not the 
$y$--dependence arising from the overall normalization, $\kappa$.
The latter is $\propto y^2$ at small $y$ but may be numerically 
important at $y \sim 1$; the curves in this region are shown
for illustrative purposes only.

In the above discussions we have considered production of a $0^+$ system.
The cross section for production of a $0^{-}$ state is significantly 
suppressed compared to $0^+$. This is because in the hard scattering process
only one gluon polarization state gives a large contribution in the 
LO approximation \cite{Gribov:2003nw}, and from one gluon polarization 
it is impossible to build a parity--conserving amplitude for the production 
of a $0^{-}$ state. 

Our discussion of the transverse momentum dependence in this Section
is based on the approximation of independent hard and soft interactions,
Eqs.~(\ref{amp_momentum}--\ref{T_rho}). One should expect that the
inclusion of correlations between hard and soft interactions,
as described in Section~\ref{sec:beyond}, would modify not
only the RGS probability but also the transverse momentum 
dependence of double--gap exclusive diffraction. This interesting
question will be addressed elsewhere.
\section{Discussion and outlook}
\label{sec:discussion}
In this paper we have outlined an approach to RGS in double--gap 
exclusive diffractive processes in $pp$ scattering based on the 
idea that hard and soft interactions are approximately independent
because they proceed over widely different time--and distance scales.
We have shown that this idea can be practically implemented in
the framework of Gribov's parton picture of high--energy scattering,
and gives rise to a conceptually clear and quantitative description
of RGS. 

In the independent interaction approximation, where correlations 
between hard and soft interactions are neglected,  
the RGS probability can be expressed in a model--independent fashion 
in terms in two phenomenological ingredients --- the gluon GPD in the 
proton, and the $pp$ elastic scattering amplitude. At this level we
recover a simple geometric picture of the interplay of hard and soft
interactions in the impact parameter representation. The fact that the 
$pp$ elastic amplitude at high energies approaches the BDL at TeV
energies suppresses small impact parameters and ensures 
dominance of peripheral scattering in double--gap diffraction.
This is crucial both for justifying the approximations made
in our derivation, and for determining the numerical value
of the RGS probability. Our numerical results for the RGS probability 
are somewhat lower than those obtained previously within the eikonalized 
Pomeron model of soft interactions; the agreement of Eq.~(\ref{S2_LHC_num}) 
with the best estimate quoted in Ref.~\cite{Kaidalov:2003fw} is accidental 
and due to the fact that these authors assume a larger radius of the 
transverse spatial distribution of gluons with $x \sim 10^{-2}$. 

Our numerical prediction for the RGS probability in double--gap exclusive 
diffractive Higgs boson production at the LHC ($m_H = 100 - 200 \, \text{GeV}, 
\sqrt{s} = 14\, \text{TeV}$) in the independent interaction approximation
is $S^2 \approx 0.03$. According to the detailed calculations of
Ref.~\cite{Kaidalov:2003fw}, 
this would put the estimated cross section for Higgs production
in the minimal supersymmetric model (MSSM) with detection through
the $b\bar b$ mode at $\text{Br}(H \rightarrow b\bar b) \, 
\sigma_{\text{diff}}(0^+) \sim \text{few} \times 10 \, \text{fb}$,
which should be accessible at the LHC; see Ref.~\cite{Kaidalov:2003fw}
for details.

Our partonic approach to RGS also allows us to discuss the effect
of correlations between hard and soft interactions on the RGS 
probability. Such correlations lower the RGS probability compared
to the independent interaction approximation. In the case of 
long--distance correlations due to the proton's pion cloud, 
we estimated this effect to be of the
order of $\sim 10\%$. Potentially more important are short--distance 
correlations, \textit{e.g.}\ those suggested by the phenomenological
concept of a ``constituent quark'' structure of the proton). 
Such correlations can be regarded as a
change of the effective size of the diffractively scattering systems,
and could reduce the predictions for the RGS probability by a 
substantial factor. In view of its importance for the Higgs boson
search at the LHC this problem clearly requires further theoretical 
study. It can in principle also be addressed experimentally, 
by ``measuring'' the RGS probability in double--gap processes for which 
the hard scattering amplitude is known, such as dijet production.

An important ingredient in our description of RGS is the 
profile function of the complex $pp$ elastic scattering amplitude. 
This underscores the importance of the planned measurements of $pp$ 
elastic scattering and total cross sections in the TOTEM experiment 
at the LHC \cite{unknown:1997xu}, 
as well as at RHIC \cite{Bultmann:2005na}. In addition to providing input
for more accurate estimates of the RGS probability in diffraction,
such measurements would allow us to further explore the fascinating
new regime of the BDL in high--energy hadron--hadron scattering.

Measurements of the transverse momentum dependence of double--gap
exclusive diffractive processes with large cross section
(dijet production) would allow one to perform detailed studies of the 
diffractive reaction mechanism. Following the strategy outlined in 
Section~\ref{sec:differential}, once the reaction mechanism has
been established, one could even use such processes to extract information
about the transverse spatial distribution of gluons in the colliding
protons, including its change with $x$. Such studies would complement
the information on the two--gluon formfactor obtained from vector meson 
production at HERA or a future electron--ion collider (EIC). Eventually, 
using QCD evolution as well as models of nucleon structure,
these data on the transverse spatial distribution of gluons could 
also be correlated with the planned measurements of quark GPDs 
in hard exclusive processes in $ep$ scattering in fixed--target
experiments (HERMES, JLab 12 GeV, COMPASS). One of the
advantages of Gribov's parton picture of hard and soft interactions
is precisely that it unifies the description of hadron--hadron
and electron/photon--hadron scattering at high energies.
Other ways to probe GPDs in $pp$ scattering with hard processes
(non-diffractive) have been described in Ref.~\cite{Frankfurt:2004kn}.

We would like to comment on some of the experimental
aspects of measurements of the transverse momentum 
dependence of double--gap exclusive diffraction with the
proposed forward detectors at the LHC. Such measurements require
good energy resolution in order to guarantee exclusivity
and determine the mass of the diffractively produced system,
as well as sufficient transverse momentum resolution to 
map the $\bm{p}_{1\perp}, \bm{p}_{2\perp}$ distributions.
An important experimental problem is that the intrinsic transverse 
momentum distribution in the beams at the interaction point (IP)
puts a lower bound on the transverse momentum transfers that can 
be resolved. This distribution is determined by the beam optics,
and thus closely correlated with the luminosity. The proposed 420 m 
forward detectors for the CMS and ATLAS experiments 
\cite{Albrow:2005ig,Cox:DIS2006} 
can tag protons in the range $0.002 \le \xi \le 0.015$; in the TOTEM 
experiment at CMS with detectors at 200 m the range will be extended to 
$\xi<0.1$ \cite{Whitmore:DIS2006}. Both detectors can obtain a longitudinal
and transverse momentum resolution comparable to the beam distributions.
At a luminosity of $10^{33} \, \text{cm}^{-2} s^{-1}$ with $\beta^* = 0.5 \, 
\text{m}$, and a one-sigma normalized emittance $\epsilon = 3.75\times 10^{-6} \, \pi \, 
m$, the one-sigma angular spread of the beams at the IP is $8 \, \mu r$, 
corresponding to a transverse momentum spread of $56 \, \text{MeV}/c$. 
This sets the scale for experimental smearing of the transverse momentum 
distributions. The TOTEM
experiment also envisages running at substantially larger $\beta^*$ values 
(18, 90 and 1540 m). These values will reduce the transverse angular spread 
of the beams at the IP by $\sqrt{\epsilon/(\pi\beta^*)}$, but with a 
concomitant reduction in luminosity. Given that the typical scale of 
the transverse momentum distributions is $m_g \approx 1\, \text{GeV}$,
it seems feasible to make detailed and precise measurements of the
transverse momentum distributions with the LHC420 and TOTEM detectors
even when running in high--luminosity mode.
\begin{acknowledgments}
We thank V.~Khoze, R.~Orava, and M.~Ryskin for useful discussions.
M.~Ryskin kindly made available to us a numerical parametrization 
of the $pp$ elastic amplitude of Ref.~\cite{Khoze:2000wk}.

Notice: Authored by Jefferson Science Associates, LLC under U.S.\ DOE
Contract No.~DE-AC05-06OR23177. The U.S.\ Government retains a
non-exclusive, paid-up, irrevocable, world-wide license to publish or
reproduce this manuscript for U.S.\ Government purposes.
Supported by other DOE contracts and the Binational Science 
Foundation (BSF).
\end{acknowledgments}
%
%

%
%
%
%
%
\end{document}